\documentclass[twocolumn]{aastex631}
\usepackage[utf8]{inputenc}
\usepackage{hyperref}
\usepackage{bm}
\hypersetup{
    colorlinks=true,
    filecolor=magenta,      
    urlcolor=blue,
}

\newcommand{\kms}{km s$^{-1}$}
\newcommand{\Mjsr}{MJy sr$^{-1}$}

\newcommand{\jybe}{Jy beam$^{-1}$}
\newcommand{\hii}{H{\scriptsize II}}
\newcommand{\hi}{H{\scriptsize I}}

\newcommand{\cms}{cm$^{-2}$}

\newcommand{\msun}{M$_{\odot}$}

\newcommand{\til}{$\sim$}
\newcommand{\ring}{M0.8--0.2}
\newcommand{\ie}{i.e.}
\newcommand{\eg}{e.g.}
\newcommand{\ratio}{\langle B_{t}^{2} \rangle/\langle B_{0}^{2} \rangle}
\newcommand{\ratiof}{\frac{\langle B_{t}^{2} \rangle}{\langle B_{0}^{2} \rangle}}
\newcommand{\dispfunct}{1 - \langle\cos[\Delta \phi(\ell)] \rangle}

\newcommand{\degree}{$^{\circ}$}

\newcommand{\ra}{$\alpha$(J2000)}
\newcommand{\dec}{$\delta$(J2000)}
\newcommand{\h}{$^{\mathrm{h}}$}
\newcommand{\m}{$^{\mathrm{m}}$}
\newcommand{\s}{$^{\mathrm{s}}$}

\begin{document}

\title{\uppercase{SOFIA/HAWC+ Far-Infrared Polarimetric Large Area CMZ Exploration (FIREPLACE) Survey II: Detection of a Magnetized Dust Ring in the Galactic Center} }

\author[0000-0002-4013-6469]{Natalie O. Butterfield}
\affiliation{National Radio Astronomy Observatory, 520 Edgemont Road, Charlottesville, VA 22903, USA}
\email{nbutterf@nrao.edu}
\author[0000-0001-8819-9648]{Jordan A. Guerra}
\affil{Department of Physics, Villanova University, 800 E. Lancaster Ave., Villanova, PA 19085, USA}
\affil{Cooperative Institute for Research in Environmental Sciences (CIRES), University of Colorado, Boulder, CO 80309, USA}
\author[0000-0003-0016-0533]{David T. Chuss}
\affil{Department of Physics, Villanova University, 800 E. Lancaster Ave., Villanova, PA 19085, USA}
\author[0000-0002-6753-2066]{Mark R. Morris}
\affil{Department of Physics and Astronomy, University of California, Los Angeles, Box 951547, Los Angeles, CA 90095, USA}
\author[0000-0002-5811-0136]{Dylan Par\'e}
\affil{Department of Physics, Villanova University, 800 E. Lancaster Ave., Villanova, PA 19085, USA}
\author[0000-0002-7567-4451]{Edward J. Wollack}
\affil{NASA Goddard Space Flight Center, Greenbelt, MD 20771, USA}
\author[0000-0002-7408-7589]{Allison H. Costa}
\affiliation{National Radio Astronomy Observatory, 520 Edgemont Road, Charlottesville, VA 22903, USA}
\author[0000-0001-9315-8437]{Matthew J. Hankins}
\affiliation{Arkansas Tech University, 215 West O Street, Russellville, AR 72801, USA}
\author[0000-0002-1414-7236]{Scott C. Mackey}
\affil{Department of Physics and Astronomy, University of California, Los Angeles, Box 951547, Los Angeles, CA 90095, USA}
\affil{Department of Physics \& Kavli Institute for Cosmological Physics, University of Chicago, Chicago, IL 60637, USA}
\author[0000-0002-8437-0433]{Johannes Staguhn}
\affiliation{Department of Physics and Astronomy, Johns Hopkins University, 3400 North Charles Street, Baltimore, MD 21218, USA}
\author[0000-0003-4821-713X]{Ellen Zweibel}
\affiliation{Department of Astronomy, U. Wisconsin-Madison, 475 N Charter Street, Madison, WI 53706, USA}


\begin{abstract}

We present the detection of a magnetized dust ring (M0.8-0.2) in the Central Molecular Zone (CMZ) of the Galactic Center. The results presented in this paper utilize the first data release (DR1) of the Far-Infrared Polarimetric Large Area CMZ Exploration (FIREPLACE) survey \citep[i.e., FIREPLACE I;][]{my23}. The FIREPLACE survey is a 214 \micron\ polarimetic survey of the Galactic Center using the SOFIA/HAWC+ telescope. The \ring\ ring is a region of gas and dust that has a circular morphology with a central depression. The dust polarization in the \ring\ ring implies a curved magnetic field that traces the ring-like structure of the cloud. We posit an interpretation in which an expanding shell compresses and concentrates the ambient gas and magnetic field. We argue that this compression results in the strengthening of the magnetic field, as we infer from the observations toward the interior of the ring.

\end{abstract}


\section{Introduction}
\label{intro}

Feedback from massive stars, in the form of strong stellar winds, radiation, and ultimately, supernovae, can inject energy and momentum into the surrounding interstellar medium (ISM). These phenomena can drive turbulence in the region, compress the surrounding gas, and propagate shocks. Such energetic sources of feedback in the CMZ can produce shell-like structures in the surrounding gas and dust \citep{tsuboi97, oka01, tsuboi09, tsuboi15, my18}. Therefore, investigating the polarization of the thermal emission from magnetically aligned dust grains in such regions can provide insight into how the feedback from massive stars has locally affected the magnetic field in the CMZ. 

\begin{figure*}
\centering
\includegraphics[width=1.0\textwidth]{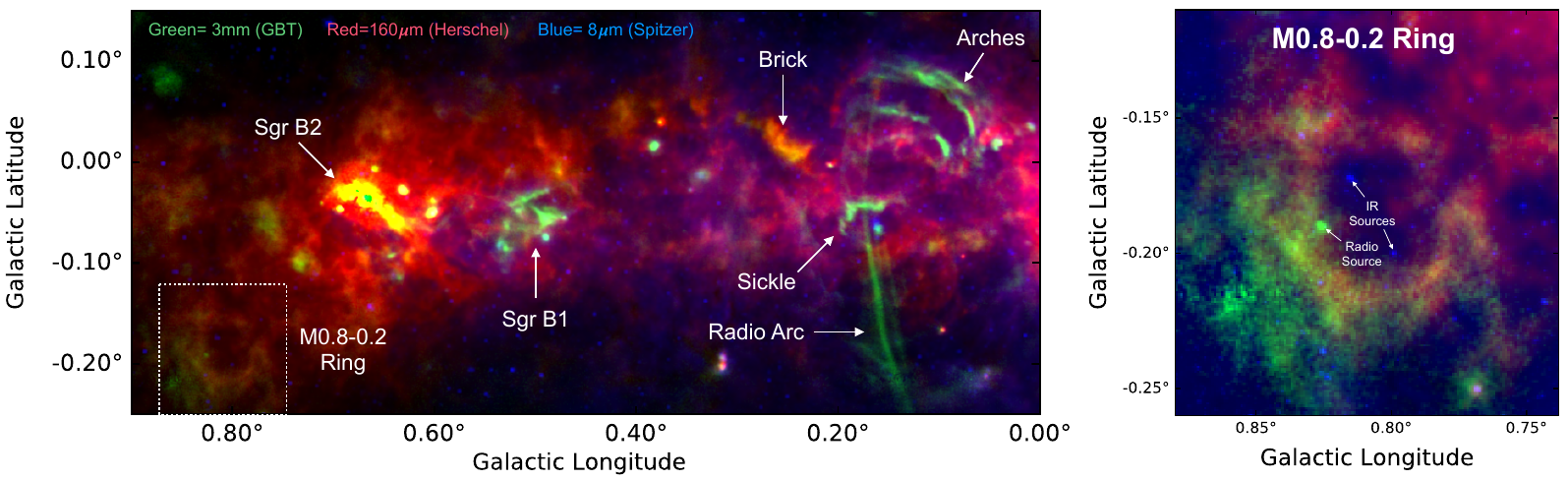}
\caption{Three-color image of the Eastern 125 parsecs of the Galactic Center (left) and the \ring~cloud (right). These images show {the GBT+MUSTANG 3 mm (90 GHz) microwave emission \citep[Mustang Galactic Plane Survey (MGPS), 9\arcsec\ angular resolution;][]{ginsburg20} in green, the far-infrared, 160 \micron\ emission from Herschel  \citep[Hi-GAL survey, PACS instrument, 12\arcsec\ angular resolution;][]{molinari16} in red,} and the 8 \micron\ emission from Spitzer \citep[GLIMPSE II survey, IRAC4 band, 1.2\arcsec\ angular resolution;][]{Churchwell09} in blue. Labeled at left are several prominent features in the CMZ. The white dashed box at the far left shows the location of the \ring\ cloud detailed in the right panel.  Labeled at right are the compact sources discussed in Section \ref{point-source}: the radio point-source (green) and the two IR point sources (blue). 
}
\label{3color}
\end{figure*}

Figure \ref{3color}, left, shows a 3 color image of the Eastern 125 pc \citep[assuming a distance of 8.2 kpc to the GC;][]{gravity} of the CMZ \citep[roughly the area covered by the DR1 FIREPLACE survey, presented in][]{my23}, {at 3 mm \citep[green; MGPS;][]{ginsburg20}, 160 \micron\ \citep[red; Hi-GAL survey;][]{molinari16}, and 8 \micron\ \citep[blue; GLIMPSE II survey;][]{Churchwell09}.} In this figure the cool dust, highlighted in red, traces the dense molecular clouds; the green emission highlights the thermal Bremsstrahlung (free-free; e.g., Sgr B2, Sickle), bright non-thermal Synchrotron emission (e.g., Radio Arc), and cool dust; and the blue color represents the PAH emission caused by the UV emission from high-mass stars. The Eastern region of the CMZ contains numerous dense molecular clouds and the well-known massive star-forming complex, Sgr B2.  Southeast of the Sgr B2 complex is the GC \ring\ cloud \citep[white box in Figure \ref{3color}, left;][]{Pierce-Price00, tsuboi15, mills17}, located \til30 pc in projection from Sgr B2.

The \ring\ structure has the morphology of a ring, with the center of the cloud showing a depression in the emission (Figure \ref{3color}, right). The \ring\ ring is roughly 4\arcmin$-$6\arcmin\ across (9$-$14 pc) and its periphery is 1\arcmin$-$2\arcmin\ thick (2$-$4.5 pc; as measured from the inner to the outer radii). As shown in Figure \ref{3color}, right, most of the cold gas and thermal radio emission (green) is concentrated toward the south and eastern region of the cloud. The warmer dust (red) is concentrated toward the north and western region of the cloud. Located within the cavity of  \ring\ are two 8 $\mu$m point-sources, shown in blue. There is also a bright radio point-source, shown in green, toward the left inner edge of the \ring\ ring. As we will show in Section \ref{point-source}, this point-source is thermal in nature \citep{heywood22} and is also detected at mid- and far-infrared wavelengths (4.5$-$70 \micron). 

The \ring\ ring was first discussed by \cite{Pierce-Price00}, in their SCUBA data (identified as PPR G0.8--0.18), who argue the source could be a wind-blown feature or a supernova remnant. Follow-up observations of the kinematics and properties of \ring\ were investigated in detail by \cite{tsuboi15} using H$^{13}$CO$^+$ (1--0) and SiO (2--1) data, obtained using the 45-m Nobeyama telescope. The kinematics presented in \cite{tsuboi15} confirmed the \cite{Pierce-Price00} hypothesis that \ring\ is an expanding shell. \cite{tsuboi15} also showed that the \ring\ ring is one of several shells they detected in this region near Sgr B2. \cite{tsuboi15} measured the kinetic energy of the expansion to be 4$\times$10$^{49}$ ergs and argued that the expansion was caused by a supernova.
They also estimate the age of the \ring\ ring to be 1.8$\times$10$^{5}$ yr and the mass to be 2.7 $\times$ 10$^3$ \msun.

The 3 mm molecular line survey of the CMZ, presented in \cite{jones12} and \cite{mills17}, shows that the cloud is rich with many molecular species that are commonly detected in CMZ clouds (e.g., HCN, HCO$^+$, etc). Additionally, several of these detected spectral lines are known to trace shocks (e.g., HNCO, SiO), indicating that \ring\ could be undergoing shocks. These shocks could be produced by the SNe interaction, as argued by \cite{tsuboi15}.

\cite{ginsburg16} conducted a 218 GHz survey of H$_2$CO across the CMZ using the 12 meter APEX telescope. They measure a gas temperature of 50$-$200 K in the cloud using the p-H$_2$CO 3$_{0,3}$--2$_{0,2}$ and 3$_{2,1}$--2$_{2,0}$ lines (see their Figure 7). This gas temperature in the \ring\ cloud is consistent with gas temperature measurements for other CMZ clouds. The gas kinematics for the \ring\ cloud, from these spectral line studies, show the cloud has a central velocity around $+$40 \kms, consistent with velocities of other clouds in the Galactic center, which can range from $-$150 to +150 \kms\ \citep[e.g.,][]{bally87, bally88, morris96, Kru15}. The similar gas temperatures and velocities indicate this cloud is likely embedded in the CMZ.

In interstellar dust clouds similar to those shown in red in Figure \ref{3color}, spinning non-spherical dust grains can become aligned with their long axis perpendicular to the local magnetic field direction in the presence of an anisotropic radiation field (i.e., $B$-RAT theory). In the far-infrared, the measurement of polarized emission from these grains can therefore be used to infer the plane-of-sky projected magnetic field direction; the direction of the inferred field in this widely-accepted hypothesis is perpendicular to the direction of the polarized electric field vector \citep[see the review on dust polarization and `$B$-RAT' theory by][]{Andersson15}.

This magnetic alignment of grains is the best-studied alignment mechanism, though there is still much work to be done for a complete understanding of dust grain physics.  Other mechanisms for aligning interstellar dust grains have been posited. These include mechanical and radiative alignment \citep{Lazarian07}. However, a recent paper by \cite{Akshaya23} investigated the dust grain alignment in the CMZ and found that the polarization observations at 214 \micron\ can best be explained using $B$-RAT theory (see their paper for a thorough investigation of grain alignment in this region).

A deep consideration of grain physics is beyond the scope of this paper. Though the various alternative alignment mechanisms are gaining attention, most of the literature to date involving polarimetric measurements of dust emission concludes that the predominant alignment direction of dust grain angular momentum corresponds to the direction of the magnetic field. This paper adopts magnetic alignment in its interpretation of the polarization data; if this paradigm shifts due to new tests of grain physics, results such as those described in this work would need to be revisited. 

In this paper we present the detection of polarized 214 \micron\ dust emission associated with the CMZ source M0.8-0.2 using data from the recent FIREPLACE survey \citep[FIREPLACE I;][]{my23}. An overview of the FIREPLACE survey is discussed briefly in Section \ref{obs}. We present the magnetic field structure and strength of the \ring\ ring in Section \ref{res} and discuss the implications of these findings in Section \ref{dis}.

\section{Observations}
\label{obs}

The results presented in this paper utilize the first data release (DR1) of a recent SOFIA/HAWC+ legacy survey of dust polarization in the CMZ:  SOFIA/HAWC+ Far-Infrared Polarimetric Large Area CMZ Emission (FIREPLACE) Survey \citep[i.e., FIREPLACE I; see][for a full discussion of the reduction methods for the data shown in this paper]{my23}. The FIREPLACE survey is a 214 \micron\ study taken with the SOFIA telescope and HAWC+ polarimeter. The complete survey \citep[FIREPLACE III;][]{Pare24} covers the inner 1.5 degree of the CMZ with a spatial resolution of 19\farcs6 (0.7 pc at the Galactic Center), comparable to the size scales of small clouds in the CMZ. For the data shown here, we used conservative SOFIA/HAWC+ threshold criteria for inclusion: polarization signal-to-noise, $p$/$\sigma_p$ $>$ 3; fractional polarization, $p$ $<$ 50\%; and signal-to-noise of the total intensity (Stokes $I$), $I$/$\sigma_I$  $>$ 200 \citep[see][for additional discussion on these selection critera]{Gordon18}.

As mentioned in the Introduction, we assume that the conditions in the \ring\ ring are such that the $B$-RAT alignment mechanism dominates. As such, in this paper, we will be considering only the inferred magnetic field pseudovectors. These are represented by rotating by 90$^\circ$ the polarization pseudovectors that are shown in \citet[][see their Figure 3]{my23}.

\section{\ring: A Magnetized Dust Ring in the CMZ}
\label{res}

\begin{figure}
\centering
\includegraphics[width=0.478\textwidth]{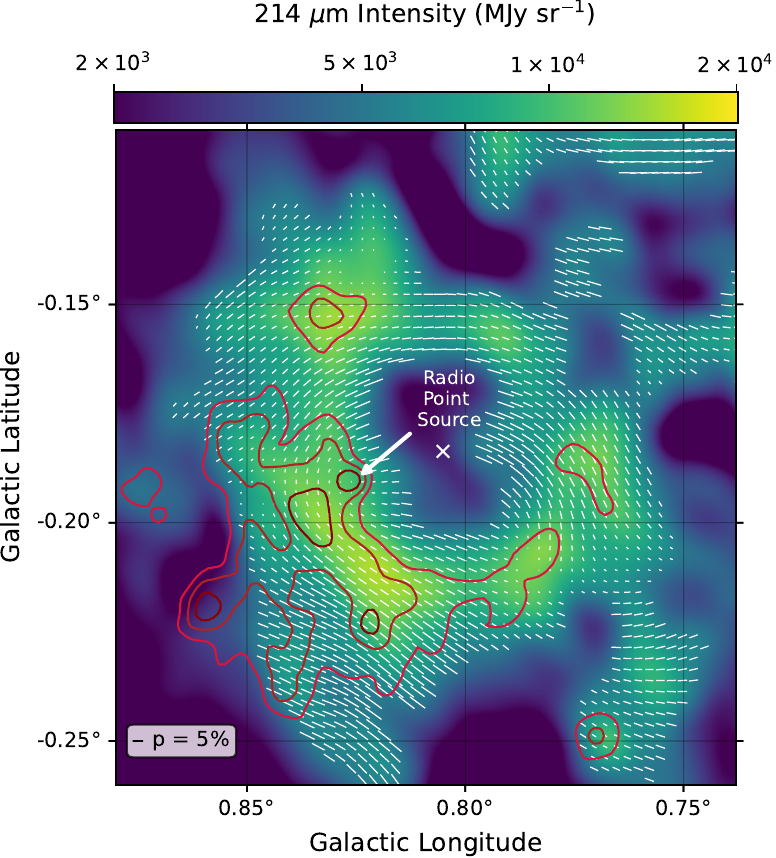}
\caption{214 \micron\ magnetic field pseudovectors in the \ring\ Ring (white pseudovectors; for $I_{214\mu\rm{m}}$$>$ 5,000 \Mjsr, spatial Nyquist sampling), overlaid on the FIREPLACE total 214 \micron\ intensity. Red contours show the smoothed 18\arcsec\ resolution MGPS data (2$\times$ the native 9\arcsec\ resolution) at 10$\sigma$, 15$\sigma$, and 20$\sigma$ \citep[for an rms value of 0.5 m\jybe;][]{ginsburg20}. This MGPS emission is shown, at the native resolution, in Figure \ref{3color} in green. The white arrow indicates the location of the radio point source, also annotated in Figure \ref{3color}. The white `$\times$' symbol marks the location of the central (r=0, $\theta$=0) position used in the polar coordinate system, shown in Figure \ref{ring-polar} and discussed in Section \ref{res}. This polar coordinate system (r,$\theta$) is centered at $l$=0\fdg8051, $b$=$-$0\fdg1836. The reference pseudovector is shown in the bottom left corner. }
\label{ring}
\end{figure}

\begin{figure}
\centering
\includegraphics[width=0.48\textwidth]{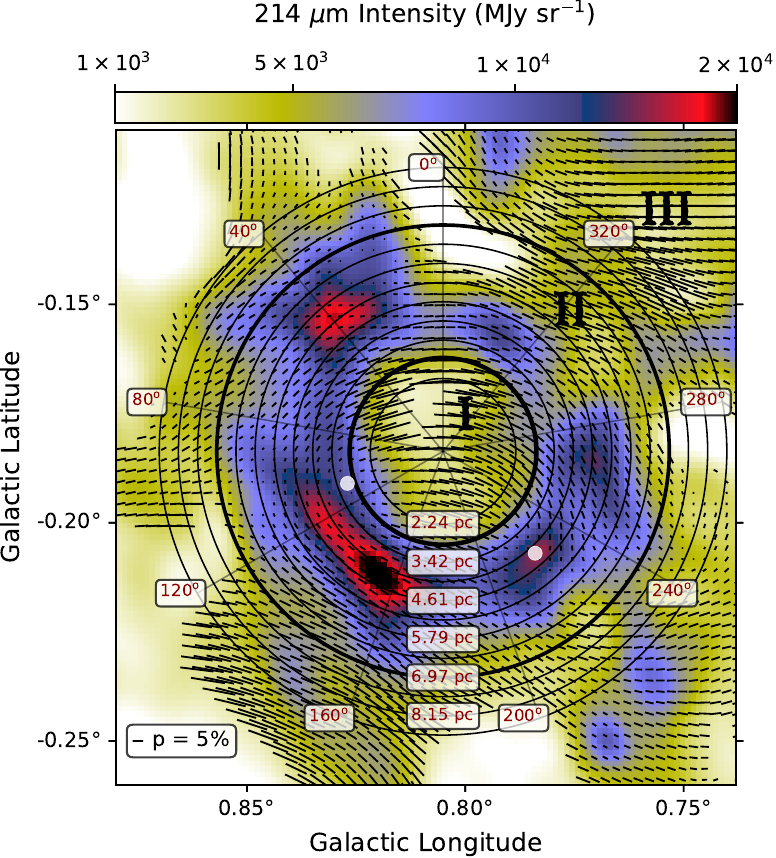}
\caption{214 \micron\ emission in the \ring\ ring. Overlaid are the 214 \micron\ magnetic field pseudovectors in the \ring\ ring, including the lower intensity pseudovectors (i.e., pseudovectors with Stokes $I_{214\mu\rm{m}}$$<$ 5,000 \Mjsr). Polar coordinate system (r,$\theta$) overlaid on the \ring\ ring, centered at $l$=0\fdg8051, $b$=$-$0\fdg1836. The $\theta$=0\degree\ direction is orientated in the direction of Galactic North and $\theta$ increases in the counter-clockwise direction, conforming to the IAU polarization standard. This polar coordinate grid is discussed in Section \ref{res} and used in Figures \ref{unwrap}--\ref{fig:disp_analsis_radii}, with the inner- and outermost radii (2.2 pc and 9.1 pc, respectively) representing the radial range for Figures \ref{fig:disp_analsis_polar} and \ref{fig:disp_analsis_radii}. The two thicker grid lines, at $r$\til3 pc and $r$\til7 pc, delineate the three different radial regions (Regions I--III, indicated by their Roman numeral) indicated in Figure \ref{unwrap}. The two white circles show the general locations of the two nodes observed in Figure \ref{unwrap}. The reference pseudovector is shown in the bottom left corner. }
\label{ring-polar}
\end{figure}

One of the more intriguing sources observed in the FIREPLACE survey is the \ring\ ring. Figure \ref{ring} shows the structure of the 214 \micron\ magnetic field pseudovectors corresponding to the brightest continuum emission in the M0.8--0.2 ring, employing the additional cut of $I_{214\mu\rm{m}}$$>$5,000 \Mjsr\ above the standard SOFIA datacuts described in Section \ref{obs}. We employ this additional cut in this figure to highlight vectors associated with the brightest emission. The observed dust polarization is the integrated polarization along the line of sight, weighted by intensity. Therefore, the lower-intensity regions are potentially more influenced by unrelated background and foreground contributions. Analysis of the polarization pseudovectors in this lower-intensity regime was conducted for the FIREPLACE I survey \citep{my23}. They found a relatively uniform distribution of the orientation angles for the field directions when compared with the higher intensity regimes, indicating that there could be source confusion due to multiple field directions along the line of sight in the higher-intensity regimes. Figure \ref{ring-polar} shows all 214 \micron\ pseudovectors from the FIREPLACE I survey that satisfy the standard SOFIA criteria, discussed in Section \ref{obs}, including the lower-intensity vectors that are associated with emission $<$5,000 \Mjsr. Most of these lower-intensity vectors can be observed towards the extreme interior and exterior peripheries of the \ring\ ring, when comparing Figures \ref{ring} and \ref{ring-polar}. 

The magnetic field has an approximately circular structure (azimuthal orientation) that generally follows the curvature of the \ring\ ring. The bright 214 \micron\ emission (i.e., $I_{214\mu\rm{m}}$$>$5,000 \Mjsr) has fractional polarization values of 1--5\%, comparable to those of other CMZ clouds in the FIREPLACE survey \citep{my23}. These fractional polarization values tend to increase towards the center of the circular structure (10--20\%; see Figure \ref{ring-polar}), suggesting grain alignment could be more efficient in this inner region. There are also four locations in \ring\ that show relatively low ($<$0.5\%) fractional polarization values: 
$l$=0\fdg784, $b$=-0\fdg207; 
$l$=0\fdg827, $b$=-0\fdg191; 
$l$=0\fdg821, $b$=-0\fdg141; and 
$l$=0\fdg845, $b$=-0\fdg193.
The low fractional polarization at these locations could be caused by a few different scenarios: 1) the field is tangled below our resolution, 2) the high density in the regions could be causing B-RAT to be less effective, or 3) there are multiple field components along our line of sight that are causing the depolarization. Since we are seeing this depolarization in small regions of the source, and not in larger, global sections of the \ring\ ring, this is likely not the case. 

\begin{figure*}
\centering
\includegraphics[width=1.0\textwidth]{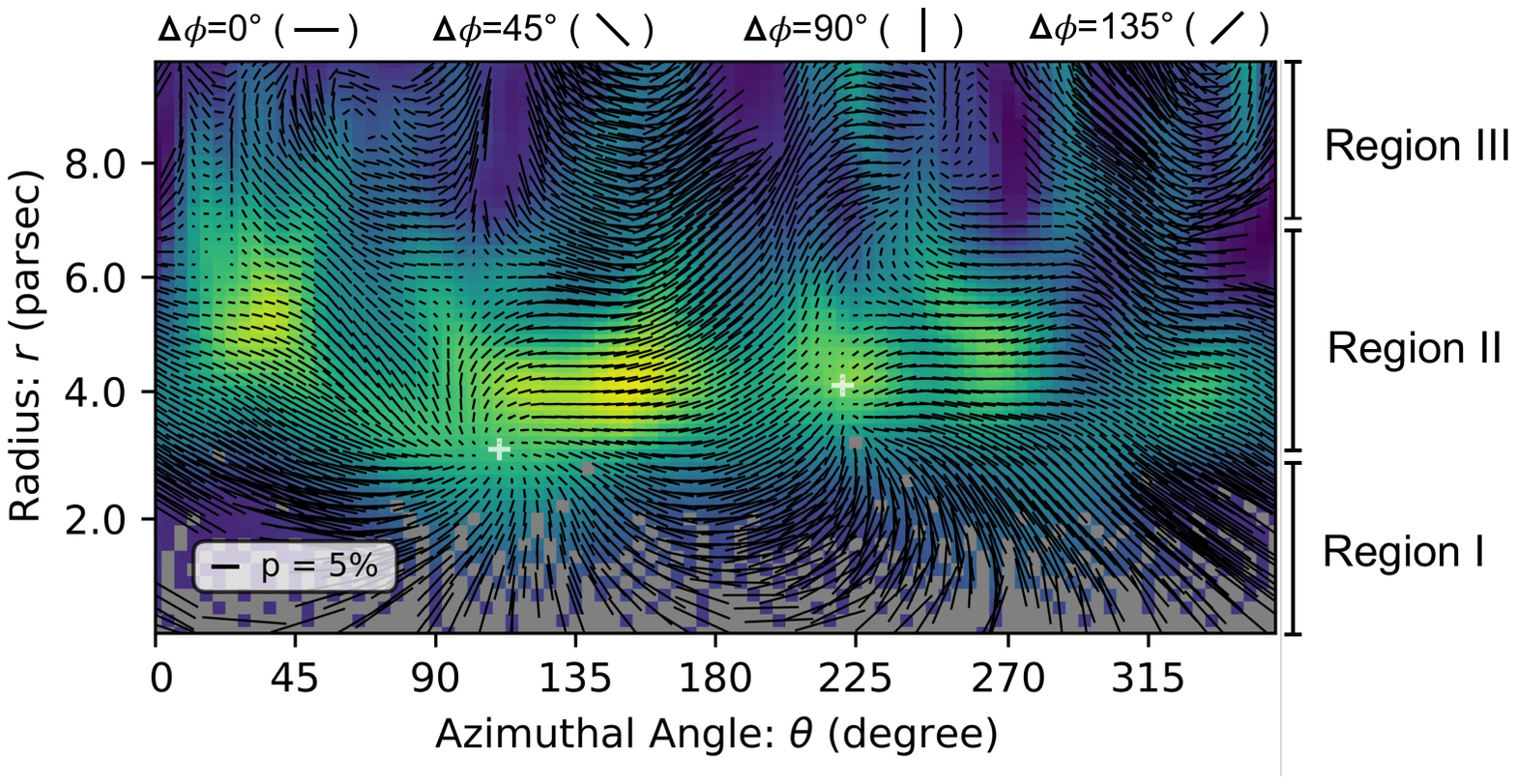}
\caption{Polarization in the \ring\ ring, from Figure \ref{ring-polar}, expressed in polar coordinates. The colorscale of the emission corresponds to the 214 \micron\ intensity (using the same scaling as shown in Figure \ref{ring}). The zero point for the polar grid is centered at $l$=0\fdg8051, $b$=$-$0\fdg1836 (white `$\times$' in Figure \ref{ring}). An azimuthal angle ($\theta$) equal to zero is oriented in the direction of Galactic North (see Figure \ref{ring}), with an azimuthal angle increasing in the counter-clockwise direction. The orientation of the $\Delta\phi$ directions, discussed in Section \ref{pol-section}, are indicated at the top of this figure for reference. The three radial zones (Regions I--III), also discussed in Section \ref{pol-section}, are annotated on the right side of the figure. The two white `+' signs show the convergence locations of the magnetic fields discussed in Section \ref{res} and marked as white circles in Figure \ref{ring-polar}.}
\label{unwrap}
\end{figure*}

The fainter 214 \micron\ emission (i.e., $I_{214\mu\rm{m}}$$<$5,000 \Mjsr), generally located towards the periphery of the cloud, shows a localized well-ordered field. For example, the southeast region of the field shows a relatively uniform magnetic field direction that is oriented roughly 20--30\degree\ to the Galactic plane. A similar magnetic field orientation is observed towards the northwest region of the field. This magnetic field orientation could be indicative of the large scale field direction detected in \cite{Mangilli19}. The northeast and southwest regions of the \ring\ ring show more variation in the magnetic field directions; however, neighboring pseudovectors are oriented in similar directions, which could be indicative of a coherent large scale field.

To better understand the circular structure of the magnetic field, we performed a coordinate transformation on the 214 \micron\ data to regrid the data from Galactic coordinates ($l$,$b$) into polar coordinates ($r$,$\theta$). Overlaid on Figure \ref{ring-polar} is this polar coordinate grid, which is centered at $l$=0\fdg8051, $b$=$-$0\fdg1836 (\ra=17\h48\m14\fs331, \dec=$-$28\degree20\arcmin35\farcs14). In the following sections we will use this polar coordinate grid to discuss the orientation of the magnetic field in relation to the structure of the \ring\ ring (Section \ref{pol-section}) and measure the magnetic field strength in \ring\ and how it varies across the entire structure (Section \ref{bfield}).

\subsection{Structure of the Dust Polarization}
\label{pol-section}

Using this new polar coordinate reference frame, shown in Figure \ref{ring-polar}, we can determine how well the magnetic field traces a ring-like structure. In an ideal case (i.e., if the magnetic field is perfectly circular) then the magnetic field orientations, expressed in polar coordinates, would be oriented perpendicular to the radial direction.

To do this coordinate transformation to the polar coordinate reference frame we re-centered the zero point of the reference frame to be at the center of the ring ($l$=0\fdg8051, $b$=$-$0\fdg1836) and averaged the data into bins of $\Delta\theta$=4\degree, and $\Delta r$=5\farcs7 (\til0.2 pc). We specified an azimuthal angle of $\theta$=0\degree\ to be oriented vertically (\ie, Galactic North) in Figure \ref{ring-polar}, conforming to the IAU polarization standard. To regrid the polarization pseudovectors, we calculated the orientation of the tangential angle ($\phi$) at each azimuthal angle ($\theta$) on the polar coordinate system. We then calculated the difference between the polarization angle (i.e., pseudovectors plotted in {Figure \ref{ring-polar}}) and this tangential angle and recorded the values as $\Delta\phi$. The resulting image for the \ring\ ring in polar coordinates is shown in Figure \ref{unwrap}. Here, a $\Delta\phi$=0\degree\ indicates the magnetic field line is aligned with the azimuthal direction and a $\Delta\phi$=90\degree\ indicates the magnetic field line is aligned with the radial direction. The orientation of the $\Delta\phi$ directions are shown at the top of Figure \ref{unwrap} for reference. 

\begin{figure}[t!]
\centering
\includegraphics[width=0.47\textwidth]{Figure5.png}
\caption{Percent polarization vs. radius in the Ring using the polar coordinate system from Figure \ref{unwrap}. The color of the data points corresponds to the 214 \micron\ intensity (shown in Figures \ref{ring} and \ref{unwrap} using the same colorscale). The vertical black lines mark the three radial regions discussed in Section \ref{pol-section} (Regions I--III). }
\label{p vs. r}
\end{figure}

We have divided this plot into three radial zones (see annotations of the right side of Figure \ref{unwrap}). We have defined Region I as the interior of the \ring\ ring ($r\lesssim$ 3 pc). Region II is defined as the emission associated with the \ring\ ring (3 pc $\lesssim r\lesssim$ 7 pc). Region III is defined as the exterior of the \ring\ ring ($r\gtrsim$ 7 pc). A significant number of vectors in Regions I and III are associated with low level emission (i.e., $I_{214\mu\rm{m}}$$<$5,000 \Mjsr). Therefore, the magnetic field orientation in these zones is likely more heavily influenced by contamination from foreground/background dust continuum. 

We find that in Region I, the field is well-organized into arc-like structures that appear to converge at two locations: $r_1$=3.2 pc, $\theta_1$=110\degree; and $r_2$=4.3 pc, $\theta_2$=220\degree. These convergence locations correspond with two of the low fractional polarization regions discussed in Section \ref{res} and can be discerned in Figure \ref{ring-polar} (\ie, white circles). Most of the emission in Region I is relatively faint ($<$5,000 \Mjsr). However, this emission shows a relatively high fractional polarization (up to 20\%) compared with other regions in \ring\ (see Figure \ref{ring-polar}). 

Region II corresponds to the brightest 214 \micron\ emission in \ring. This emission appears clumpy in the polar coordinate system, containing at least 5 main clumps around the structure. These clumps are also visible in Figure \ref{ring-polar}. The two clumps on the eastern side ($\theta$=0--180\degree) are brighter and more extended than the three clumps on the western side ($\theta$=180--360\degree). In Region II we find that the field is also organized, containing large regions of coherent emission that have magnetic fields oriented perpendicular to the radial direction of the ring (\ie, $\Delta\phi$$\simeq$0\degree). Most of these regions that show a horizontal magnetic field orientation in Figure \ref{unwrap} are associated with the brightest 214 \micron\ emission (\eg, $\theta$$\simeq$135\degree\ $r$$\simeq$5--6 pc). We note, however, that there are locations in Region II where we observe $\Delta\phi\neq0$. For example, at an azimuthal angle of $\theta$\til70\degree\ we observe a tilt in the magnetic field of roughly $\Delta\phi$$\simeq$45\degree. There are roughly four sections in Figure \ref{unwrap} which show this deviation: $\theta$\til0\degree, 70\degree, 180\degree, and 310\degree\ (see Figure \ref{unwrap}). Most of these deviating regions are between the bright clumps. We also note that in Region II there is one azimuthal angle where the magnetic field is oriented in the radial direction (i.e., $\Delta\phi$$\simeq$90\degree): $r$\til4 pc, $\theta$\til100\degree.

In Region III the field is less coherent than that observed at smaller radii ($r\lesssim7$ pc). While individual portions of Region III appear to have their own organized field shapes, as a whole, the larger-radii regions do not appear to be organized in the same circular pattern as observed at smaller radii. When observing the magnetic field orientation at radii larger than \til7 pc (see Figure \ref{ring-polar}), the field appears to be well organized in a manner consistent with the local field orientation surrounding \ring. Therefore, we can conclude that the ring-like structure in the magnetic field appears to decrease to value around 0.5 mG at radii larger than \til7 pc, as indicated in Figure \ref{unwrap}.

\begin{figure*}[t!]
\centering
\includegraphics[width=0.55\textwidth]{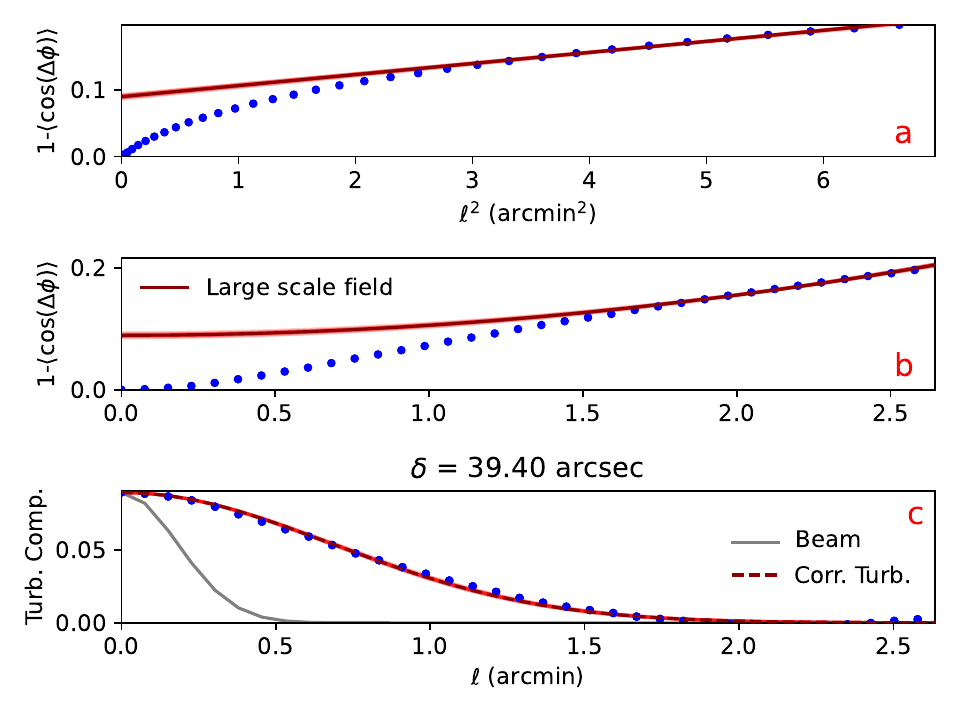}~
\includegraphics[width=0.415\textwidth]{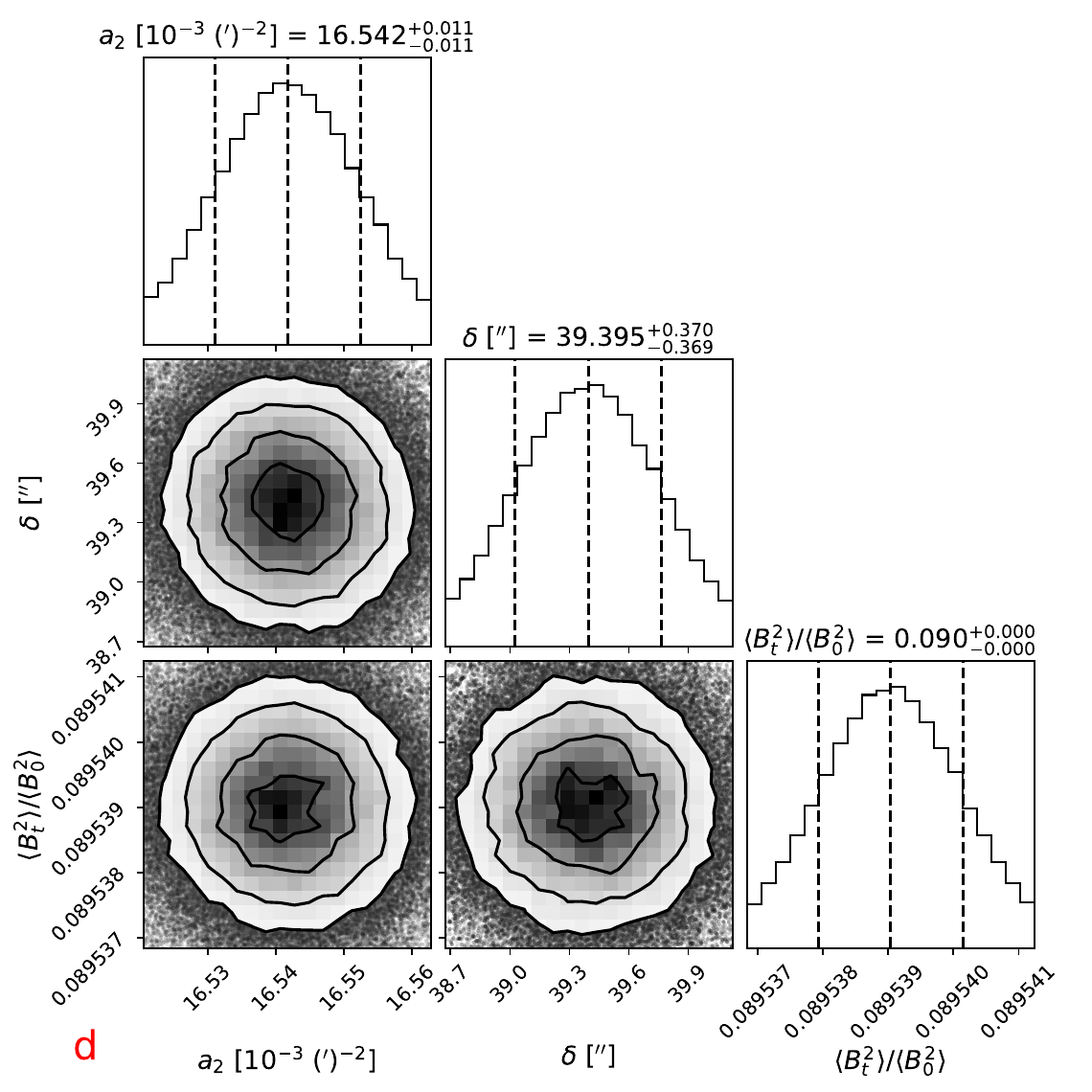}
\caption{Dispersion analysis results: a) two-point dispersion as a function of $\ell^{2}$ calculated for the entire \ring\ ring, out to a radius of \til9 pc (blue circles). The red solid line correspond to the best fit for the large-scale contribution -- second term in Eq. \ref{eq:disp_funct}. b) same as a) but as a function of $\ell$. c) auto-correlated turbulent component (blue circles). This part of the dispersion function is described by the first term in Eq. \ref{eq:disp_funct}. The best-fitting model is shown with a solid red line in panels a-c. The grey line in panel c corresponds to the auto-correlated beam profile. The plots on the right, panel d, present the posterior distributions for the best-fit parameters obtained with the MCMC solver.
}
\label{fig:disp_analsis}
\end{figure*}

Using the coordinate system shown in Figure \ref{unwrap}, we plot the percent polarization values for each data point at it's corresponding radii (Figure \ref{p vs. r}). This plot illustrates a clear trend in the distribution of polarization values as a function of radius. Here we observe that the polarization is lowest in Region II when compared with Regions I and III. Furthermore, we note a steady decrease in the polarization from the innermost radii to a radius of \til4 pc where all polarization values are $\leq$5\%. At larger radii, the polarization appears to show a wider distribution in value, ranging from 1--17.5\%. 

\subsection{Magnetic Field Strength in \ring}
\label{bfield}

The strength of the plane-of-sky (POS) component of the magnetic field in the polarized \ring\ cloud can be estimated by using the Davis-Chandrasekhar-Fermi \citep[DCF;][]{Davis1951,Chandrasekhar1953} method. The DCF relies on three measurements: mass density ($\rho_m$), velocity dispersion of the gas ($\sigma_{v}$), and the angular dispersion of the polarization vectors ($\sigma_{\phi}$). 
These three variables are combined according to the following equation for the magnetic field strength in the plane of the sky:
\begin{equation} 
B_{\rm POS} = \sqrt{4\pi \rho_m} \frac{\sigma_v}{\sigma_\phi}.
\label{eq:bfield eq}
\end{equation}

Typically, $\rho_m$ values are estimated from the column density while $\sigma_{v}$ is inferred from the linewidth of a molecular tracer, assuming isotropic turbulence.

The angular dispersion, $\sigma_{\phi}$, on the other hand, can be estimated in different ways. We follow the angular dispersion analysis developed by \citet{Hildebrand2009} and \citet{Houde2009} which has been successfully applied to HAWC+ data \citep{Chuss2019,Lopez_Rodriguez2021,Guerra2021}. This approach allows a more appropriate estimation of $\sigma_{\phi}$ by separating the contribution to the measured dispersion attributable to the geometry of the large-scale field from the contribution arising from the small-scale turbulent motions. In this way, values of $B_{\rm POS}$ calculated with Eq.~\ref{eq:bfield eq} are accurate (within the limitations of the DCF approximations; see discussion below) and no correction factor is needed \citep[e.g.,][]{Ostriker2001}. More formally, the angular dispersion is quantified by the structure function $\dispfunct$, which accounts for the difference $\Delta\phi$ between all pairs of $\phi$ separated by a distance $\ell$. This dispersion function can be modelled as

\begin{equation}
    \dispfunct = \frac{1-e^{ -\ell^{2}/2(\delta^{2} + 2W^{2})}}{1 + \mathcal{N}\left[\frac{\langle B_{t}^{2}\rangle}{\langle B_{0}^{2}\rangle}\right]^{-1}} + a_{2}\ell^{2},
    \label{eq:disp_funct}
\end{equation}

\noindent
where the second term on the right describes the large-scale contribution while the first term describes the small-scale, Gaussian turbulent component which accounts for the line-of-sight and in-beam signal integration. In Eq.\ \ref{eq:disp_funct}, $\delta$, $W$, $\ratio$ correspond respectively to the turbulence correlation length, the observing beam's radius, and the ratio of turbulent-to-ordered magnetic energy. $\mathcal{N}$, where $\mathcal{N}=\Delta^{\prime}(\delta^{2}+2W^{2})/\sqrt{2\pi\delta^{3}}$, is the number of turbulent cells along the line of sight. The cloud's effective depth, $\Delta^{\prime}$, is estimated from the width of the auto-correlation function of the polarized flux, assuming that the distribution of the gas density along the LOS is similar to that in the POS.

\begin{figure}
\centering
\includegraphics[width=0.46\textwidth]{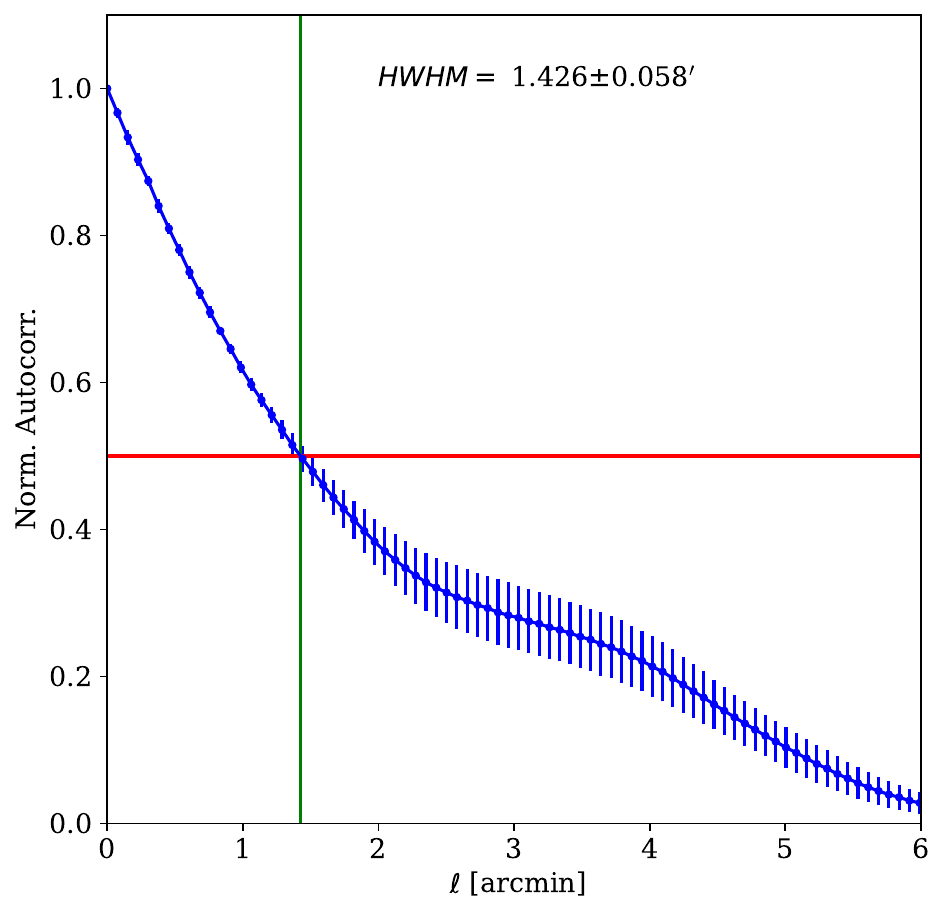}
\caption{The \ring\ ring's auto-correlation function. This one-dimensional, isotropic function (blue symbols) is calculated from the two-dimensional auto-correlation of the polarized flux, by averaging in the azimuthal direction. Blue bars corresponds to the standard error for average value. The half-width half-max distance (where the vertical green line intersects the horizontal red line) corresponds to the estimation of $\Delta^{\prime}=1.43\pm0.06$\arcmin.
}
\label{fig:autocorr_plot}
\end{figure}

First, we apply this dispersion analysis to the polarization data for the entire ring structure, excluding those pixels with intensity $<$5000 \Mjsr\ and limiting the analysis to a radial distance $\sim$ 9 pc from the ring center position. The result from this analysis are shown in Figures \ref{fig:disp_analsis} and \ref{fig:autocorr_plot}. Constructed dispersion function is shown in the {\it Left} panels of Figure \ref{fig:disp_analsis} (panels a-c), while the {\it Right} panel (panel d) shows the corner plot of the best-fit parameter for Eq. \ref{eq:disp_funct}. In Figure \ref{fig:disp_analsis}a, blue circles correspond to the dispersion as a function of $\ell^{2}$. The red solid line shows the best-fit large-scale contribution to the dispersion -- the term $\propto \ell^{2}$ in Eq. \ref{eq:disp_funct}. Figure \ref{fig:disp_analsis}b is very similar to Figure \ref{fig:disp_analsis}a but as a function of $\ell$ rather than $\ell^{2}$. The relevance of panel \ref{fig:disp_analsis}a lies in clearly identifying the best range of $\ell$ values that corresponds to the large-scale field contribution since in this version of the plot such a range corresponds to the region of the dispersion function that is clearly linear. For this dispersion function, that range is identified for $\ell \gtrsim 2^{\prime}$. 

Figure \ref{fig:disp_analsis}c displays the auto-correlated turbulent part of the dispersion function. That is, the remainder of the dispersion function at small-scales ($\ell < 2^{\prime}$) arises from turbulent motions of the magnetic field. This contribution is modelled by the first term in Eq. \ref{eq:disp_funct}. Blue circles correspond to the measured dispersion function while the red line is the best-fit model. Fitting of the dispersion function was performed through a Markov-chain Monte Carlo (MCMC) approach using the python package \texttt{emcee} \citep{Foreman-Mackey2013}.
In Figure \ref{fig:disp_analsis}c it is clear that the auto-correlated turbulent function is wider than the auto-correlated beam (grey solid line). Because $\delta > \sqrt{2}W$, the polarimetric observations resolve the turbulence and the dispersion is not underestimated by sub-beam structure. In this case, $\delta$ = 39\farcs39 $>$ 10\farcs90 (Figure \ref{fig:disp_analsis}d).

While $\delta$ can be determined directly from the fit of Eq. \ref{eq:disp_funct}, in order to determine the value of $\ratio$, we need to also estimate the value of $\Delta^{\prime}$. This value is obtained from the width of the autocorrelation function of the polarized flux in ring (Figure \ref{fig:autocorr_plot}). In this Figure we see that the half-width half-maximum value correspond to 1.43$\pm$0.06$^{\prime}$, or 3.44$\pm$0.14 pc. This depth value appears consistent with the observed thickness of the ring in 214 \micron\ intensity. Using this value of $\Delta^{\prime}$, we determined $\ratio =$ 0.090$\pm$0.001. Since $\ratio<1$, the energy corresponding to the large-scale, ordered magnetic field is larger than that of the turbulent magnetic field. In fact, from the value of $\ratio$, we can estimate that the LOS-averaged magnitude of the turbulent component is $\sim$1/3 the magnitude of the large-scale, ordered magnetic field. This supports the interpretation of a clear ring-like structure of the magnetic field with local deviations due to turbulence in the gas.

According to this dispersion analysis, to calculate the values of the POS magnetic field strength, the DCF approximation can be re-expressed as

\begin{equation} 
B_{\rm POS} = \sqrt{4\pi \rho_m} \sigma_v \left[ \ratiof \right]^{-1/2}.
\label{eq:bfield eq_1}
\end{equation}

\begin{figure*}
\centering
\includegraphics[width=0.49\textwidth]{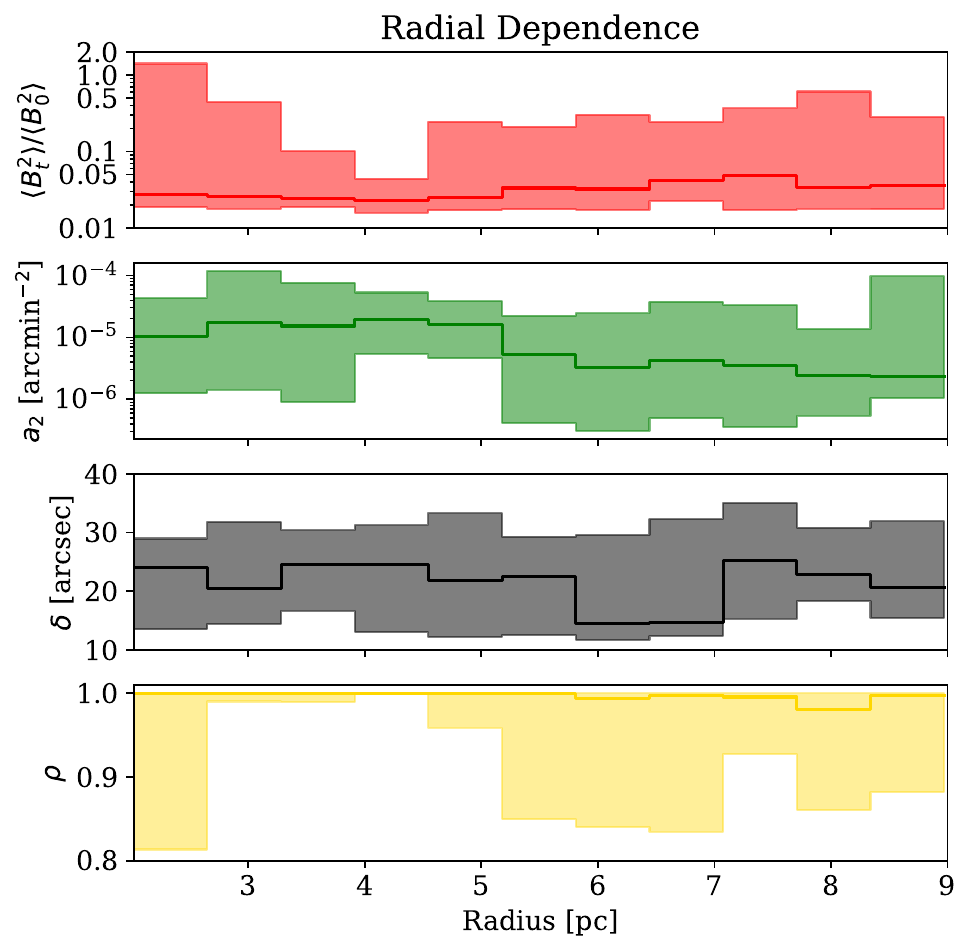}
\includegraphics[width=0.5\textwidth]{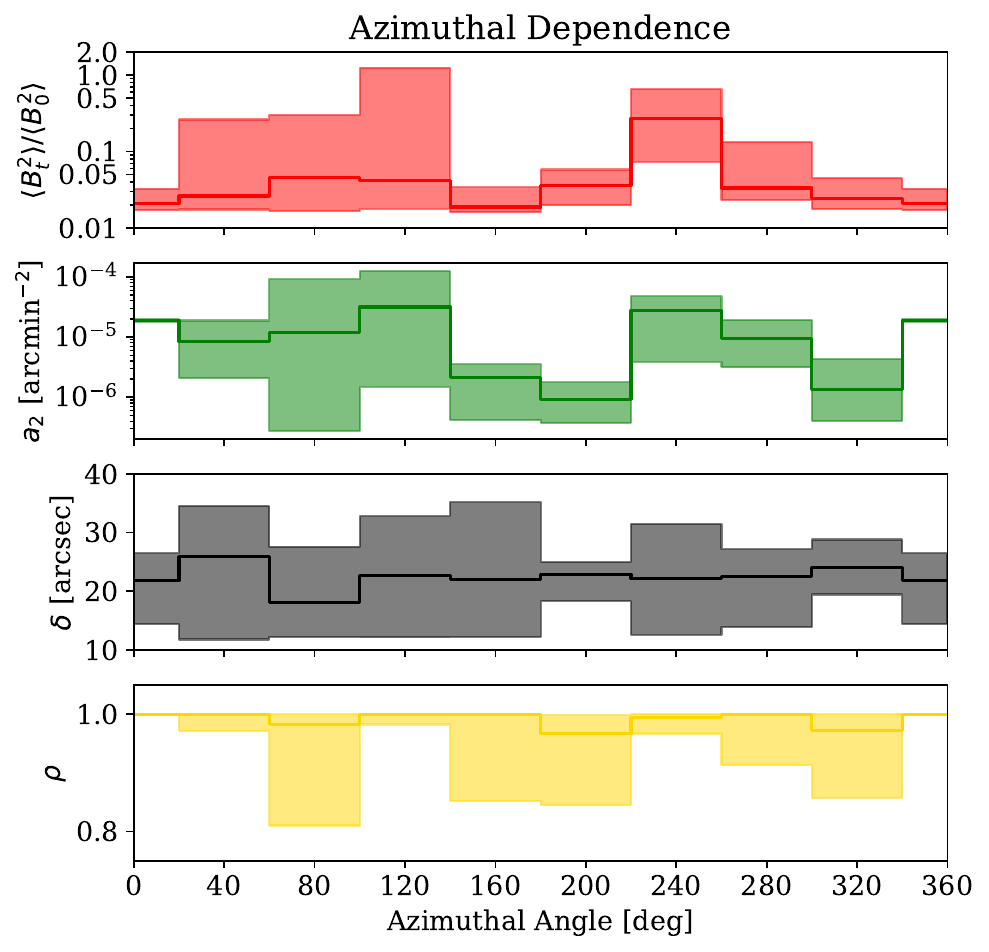}
\caption{\ring\ Radial ({\it Left}) and azimuthal ({\it Right}) profiles for the parameters derived from the polarization-angle dispersion analysis (from top to bottom): $\ratio$, $a_{2}$, $\delta$, $\rho$. Solid lines correspond to median values and the shaded area around them represents the 5th (lower edge) and 95th (top edge) percentiles.}
\label{fig:disp_analsis_polar}
\end{figure*}

To evaluate this expression for the entire ring, mass density is calculated as $\rho_m = \mu m_{H} n(H_{2})$, where $\mu(=2.8)$ is the mean molecular weight, $m_{H}(=1.67\times10^{-24}$g) is the hydrogen mass, and the number density $n(H_{2})=N(H_{2})/L$ is calculated from the average column density map presented in \citet{Marsh2015,Marsh2017} assuming a cloud depth $L =$ 1.06$\pm$0.04$\times$10$^{19}$ cm (or 3.44$\pm$0.14 pc, corresponding to the value of $\Delta^{\prime}=$ 1.43$\pm$0.06$^{\prime}$ at a distance of 8.3 kpc). The average column density is calculated over the same group of pixels considered for the dispersion analysis. We obtain $N(H_{2})=(1.46\pm0.01)\times10^{23}$cm$^{-2}$. 

The velocity dispersion value is calculated from previous HNCO (4--3) observations \citep[][further analysis of the HNCO kinematics is discussed in Section \ref{shell}]{jones12}. First, we calculate the second spectral moment map of the HNCO emission and then transform it to dispersion by taking its square root. Before applying the pixel cut, it is necessary to adjust the $\sigma_{v}$ values to account for the differences in angular resolution between the HNCO and HAWC+ measurements. Thus, according to the Larson relation \citep[\eg,][]{Larson81,Heyer2009}, we multiply the $\sigma_{v}$ map by the factor $L=(W_{\rm HAWC+,214\micron}/W_{\rm Mopra,3mm})^{1/2}$, where $W_{\rm HAWC+,214\micron}$ and $W_{\rm Mopra,3mm}$ are the beam sizes of the HAWC+ 214 $\micron$ and Mopra 3 mm data, respectively. We thereby find the average velocity dispersion for the DCF analysis to be $\sigma_{v}=1.91\pm0.01$ \kms. Using all the values above, and Eq. \ref{eq:bfield eq_1}, we calculate $B_{\rm POS}=572\pm12$ $\mu$G.\footnote{Uncertainties reported on $B_{\rm POS}$ values correspond to the propagation of variables' uncertainties through the DCF expression and it does not indicate the systematic error associated with the use of the DCF approximation, which is typically 20-70\% depending on the turbulence regime \citep[see discussion in][]{Skalidis2021}.}

Equation \ref{eq:bfield eq} above, known as the ``classical" DCF approximation, is derived from the assumption of Alfv\'enic turbulence alone in the gas. Under this assumption the dominant energy term to balance the kinetic energy perturbation is due to magnetic field fluctuations. Thus, using the classical DCF often results in a mis-estimation of the POS magnetic field strength in clouds where compressional turbulence might be important \citep[i.e.,][]{Skalidis2021} or where large-scale and/or sheared flows are prominent \citep[i.e.,][]{2023AJ....166...37G,Lopez_Rodriguez2021}. Clouds in the Galactic center have been reported to have incompressible turbulent motions, according to measured scaling exponents of power spectra calculated using velocity measurements of \hi\ emission, \hi\ absorption, and CO emission \citep{Elmegreen2004}. However, the particular conditions of the \ring\ ring, specifically the possibility of interaction with shock fronts, suggest that in some regions of the \ring~ring, compressible turbulence might develop \citep[see e.g.][]{Inoue2009}. In such cases, the use of the compressional DCF approximation might be applicable. 

The compressional-DCF POS magnetic field strength ($B_{\rm POS}^{C}$) can be calculated from the classical-DCF approximation ($B_{\rm POS}$) as

\begin{equation}
    B_{\rm POS}^{C} = \sqrt{0.5\sigma_{\phi}}B_{\rm POS}.
\end{equation}

Which, using the values from above, results in $B_{\rm POS}^{C} = 221\pm5$ $\mu$G. Thus, the most likely POS magnetic field strength for the \ring~ring is $572\pm12$ $\mu$G with values of $221\pm5$ $\mu$G in some locations. These two $B_{\rm POS}$ values we are reporting strongly depend on which physical conditions of the turbulence are operating at the point of interest. Further observations are necessary to determine whether and where shocks are present within the cloud, thereby highlighting the regions where this compressional-DCF method might best be applied.

\begin{figure*}
\centering
\includegraphics[width=0.49\textwidth]{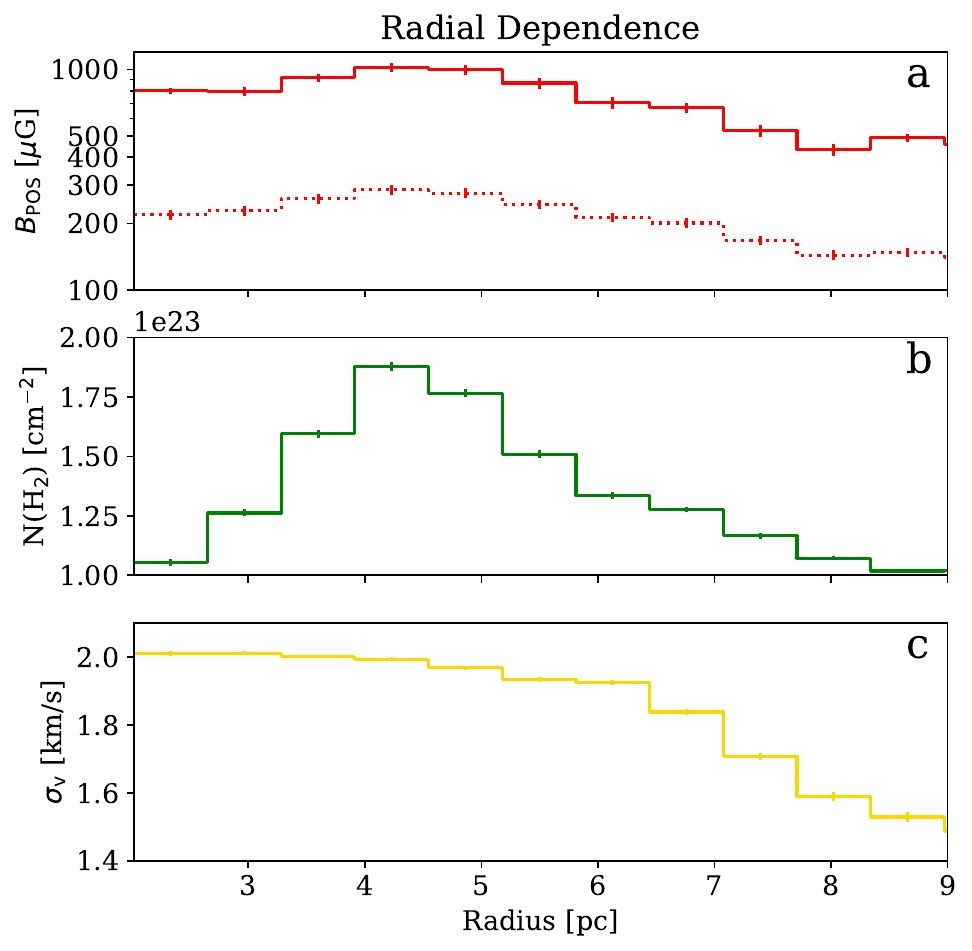}
\includegraphics[width=0.5\textwidth]{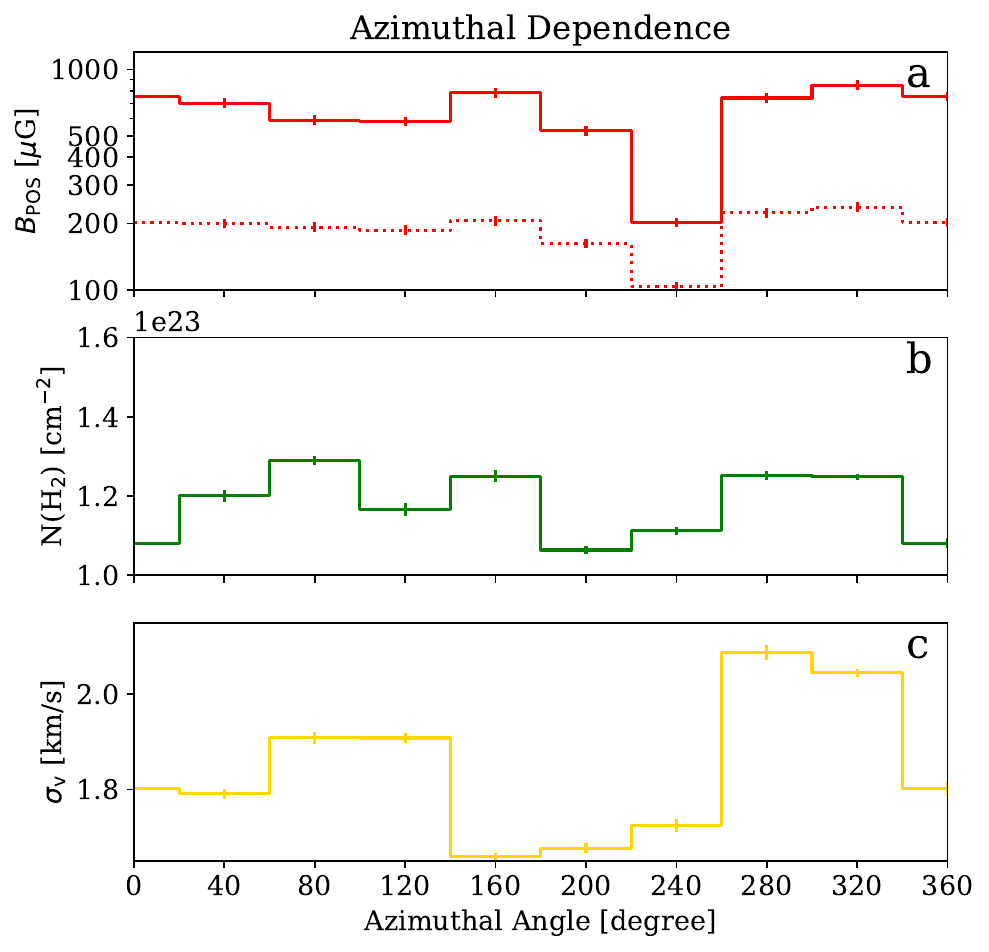}
\caption{
{\it Left:} Radial profiles of \ring: a) POS magnetic field strength, b) column density, and c) gas velocity dispersion. Values of $\sigma_{v}$ and $N(H_{2})$ correspond to averages calculated over concentric annuli centered at ($l$=0\fdg8051, $b$=$-$0\fdg1836). {\it Right:} Angular profiles of \ring; similar to ({\it Left}) but for angular wedges measured east of north, as shown in Figure \ref{ring-polar}. In both the left and right panels of (a), the solid and dotted lines correspond to the POS magnetic field strength calculated using the classical and compressional DCF approximations, respectively.} 
\label{fig:disp_analsis_radii}
\end{figure*}

An important caveat to mention at this point is the importance of properly determining the dispersion due to the large-scale, ordered magnetic field on the value of $B_{\rm POS}$. As shown in Figure \ref{fig:disp_analsis}a--b, the best-fit for the large-scale contribution determines the overall shape of the autocorrelated turbulent part (Figure \ref{fig:disp_analsis}c). Such large-scale contribution is described, according to Eq. \ref{eq:disp_funct}, by the term $\propto\ell^{2}$. This approximation appears to work well for large-scale, ordered fields that do not display a specific geometry. Thus, in the case of \ring, this approximation might not be sufficient because of the ring-like shape - which could result in the model mis-identifying some structure as turbulent dispersion, thus resulting in an underestimate of the magnetic field strength.
Therefore, performing the dispersion analysis in `wedges' ($\Delta$r,$\Delta\theta$), where the large-scale contribution is better described by the $\ell^{2}$ term, might be more appropriate and allow for more accurate values of $B_{\rm POS}$ to be determined.

Motivated by the argument exposed above, we examine variations of $B_{\rm POS}$ in both the radial and azimuthal directions. To do so, we first performed the dispersion analysis for the polar grid shown in Figure \ref{ring-polar}. That is, for a particular value of ($r,\theta$), a dispersion function is calculated for all pseudovectors in the range $r-\Delta r/2 <r\leq r+\Delta r/2$ and $\theta-\Delta\theta/2 <\theta \leq \theta+\Delta\theta/2$; where $\Delta r$ is 0.63 pc and $\Delta\theta$ is 40$^{\circ}$.\footnote{These values of $\Delta r$ and $\Delta\theta$ were chosen to guarantee that enough pairs of pseudovectors are available inside the analysis sub-regions for constructing a dispersion function with a good fit. Smaller values were tested, however they resulted in many dispersion functions with poor fits. See \citet{Guerra2021} for a similar analysis.} Each dispersion function was then fitted with the model of Eq. \ref{eq:disp_funct} assuming the same depth value ($\Delta^{\prime}$) as used above. The results of this analysis are parameters in two-dimensional, polar coordinates: $\ratio(r,\theta)$, $a_{2}(r,\theta)$, $\delta(r,\theta)$ (in addition to $\rho(r,\theta)$, the nonlinear correlation coefficient, which measures the goodness of fit). Here we present these results separately by averaging over one of the coordinates. Figure \ref{fig:disp_analsis_polar} presents the radial ({\it Left}) and azimuthal ({\it Right}) profiles of the parameters derived from the dispersion analysis. On each side of Figure \ref{fig:disp_analsis_polar}, four panels are shown (from top to bottom):  $\ratio$, $a_{2}$, $\delta$, and $\rho$. Step-like curves in these panels correspond to median values while the shaded area represents the 5th (lower edge) and 95th (upper edge) percentiles of their distributions. Only fits with $\rho>0.7$ were considered for these profiles.
We see in the profiles of Figure \ref{fig:disp_analsis_polar} that there is more variation of the dispersion analysis parameters with the azimuthal coordinate \textbf{than} with the radial one. In both cases, $\rho$ values (bottom panels) are very close to one on average, implying that our fits describe well the dispersion function and thus the determined parameters are more accurate. Moreover, values of $\delta$ exceed $\sqrt{2}W$ = 10\farcs90, which implies that the polarization angle dispersion is well-constrained for all radii and azimuths. The radial profile of $\ratio$ shows relatively constant values across the entire ring (Figure \ref{fig:disp_analsis_polar}, left top panel). However, for the majority of radial values, the spread of values can approach unity. On the other hand, the azimuthal profile of $\ratio$ shows relatively low values for most angles across the ring, with the exception of $\theta$\til240\degree, where the ratio of turbulent-to-ordered magnetic energy ($\ratio$) increases by an order of magnitude compared with neighboring azimuthal bins (Figure \ref{fig:disp_analsis_polar}, right top panel).

Figure \ref{fig:disp_analsis_radii} presents the radial (\textit{Left}) and azimuthal (\textit{Right}) profiles of the plane-of-sky magnetic field ($B_{\rm POS}$, a), column density ($N(H_{2})$, b), and velocity dispersion ($\sigma_{v}$, c) used in the DCF calculation (see Eq. \ref{eq:bfield eq_1}, along with the $\ratio$ profile shown in Figure \ref{fig:disp_analsis_polar}). Two values of $B_{\rm POS}$ are presented in both the left and right panels of (a) of Figure \ref{fig:disp_analsis_radii}: the solid line corresponds to the classical DCF approximation while the dotted line corresponds to the compressional DCF approximation. The column density radial profile shows a structure consistent with a ring of gas and dust, which peaks around $r$\til4 pc, while the velocity dispersion profile shows a constant value $\sim$2.0 \kms\ up to $r\sim$ 6 pc, before gradually decreasing to a value of 1.5 \kms. Both $B_{\rm POS}$ profiles display very similar behavior with radius; an overall decreasing tendency with increasing radius: $B_{\rm POS}(r\approx 2$ pc$)/B_{\rm POS}(r\approx 9$ pc$) = 1.75$ ($1.55$ for dotted line).  However, we note a slight increase from \til750 $\mu$G (\til220 $\mu$G) at the inner radii (2--3 pc) to \til1000 $\mu$G (\til300 $\mu$G) at $r=$ 4--5 pc. For larger radii, the magnetic field strength decreases to \til500 $\mu$G (\til140 $\mu$G) at \til8--9 pc.

On the other hand, both azimuthal profiles (Figure \ref{fig:disp_analsis_radii}, {\it Right}) show some variation with respect to the azimuthal angle. The $B_{\rm POS}$ profile shows a fairly consistent magnetic field strength -- with typical values around 500--750 $\mu$G (160--205 $\mu$G) -- with the exception of around $\theta$\til240\degree, where we measure a decrease in the field strength to \til250 $\mu$G (\til100 $\mu$G). This decrease is produced by the increase in the $\ratio$ value observed at that azimuthal angle. The velocity dispersion also shows some variation in the azimuthal direction showing two main peaks around $\theta$\til100\degree\ and $\theta$\til300\degree. These peaks in the velocity dispersion will be discussed again briefly in Section \ref{shell}.
The column density values do not show much variation, ranging from 1.0--1.5 $\times$10$^{23}$ \cms\ across all azimuthal angles. 

One of the most prominent sources of error in the calculated $B_{\rm POS}$ values is the assumed depth of the cloud. In evaluating the DCF expression, we assume a constant depth for all radii and azimuthal angles. If the ring structure is almost entirely in the POS, we can expect a depth profile that slowly decreases away from the ring's radial midpoint. This varying values of depth will affect the measured $\ratio$ and mass density but the resulting $B_{\rm POS}$ might not be much different than those here presented. However, if the observed ring corresponds to the two-dimensional projection of a spherical dust shell, then the depth radial profile should decrease from the inner to the outer radius. In such a case, the resulting radial profile of $B_{\rm POS}$ might be different from the one here presented. Therefore, a more thorough determination of the geometry in \ring\ is needed to provide a more realistic radial profiles of the magnetic field strength. 

\begin{figure*}
\centering
\includegraphics[width=0.99\textwidth]{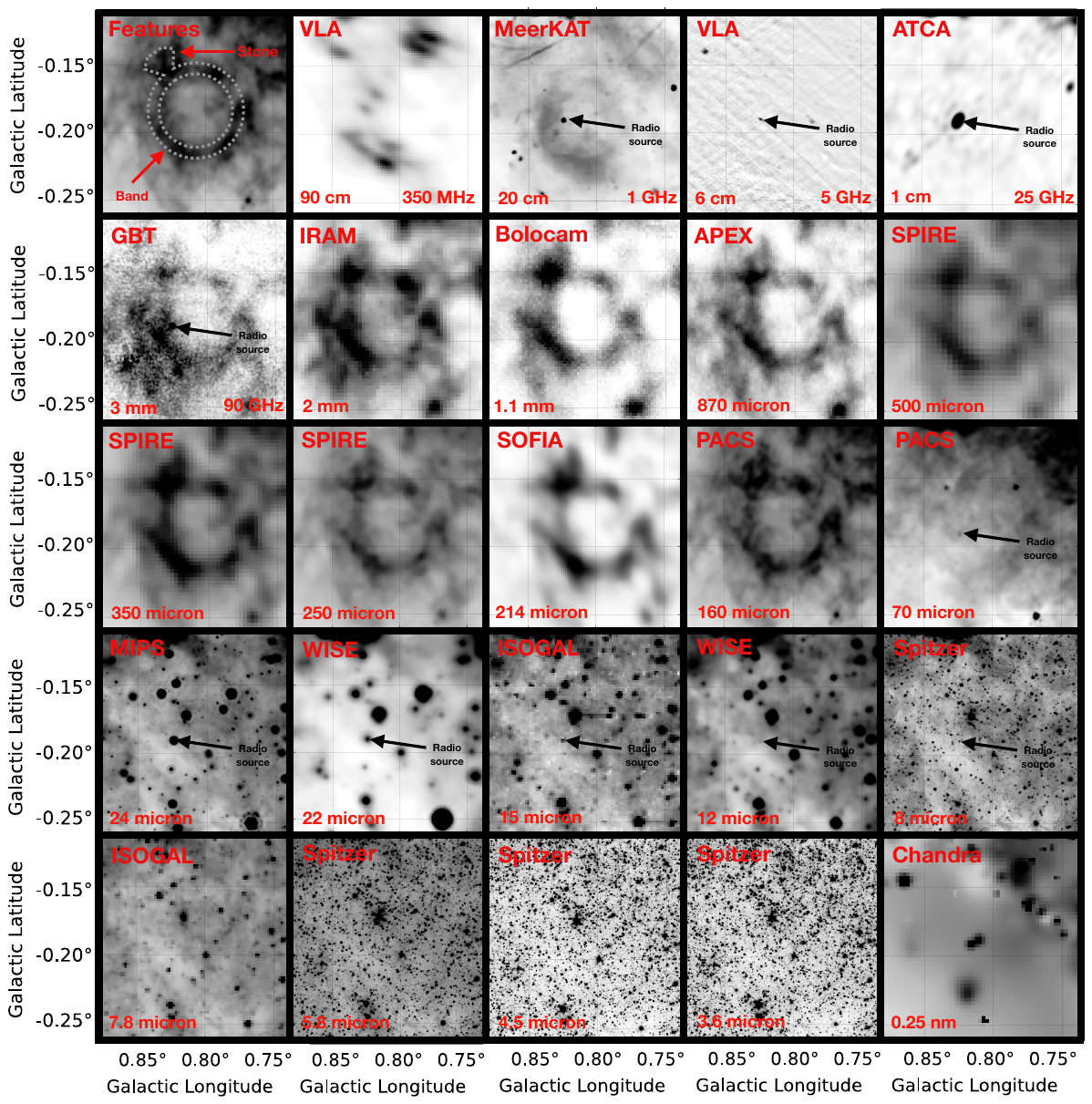}
\caption{
Emission toward the \ring\ ring from 90 cm (350 MHz) to 0.25 nm (1200 PHz; {5 keV}). The imaging parameters and citations for the data shown here are included in Table \ref{Images}. The top left panel, labeled as `Features', shows the PACS 160 micron image with annotated features that are discussed in Section \ref{multiwave}. 
}
\label{panel}
\end{figure*}

\section{Discussion}
\label{dis}

In the following sections we compare the 214 \micron\ observations from the FIREPLACE survey to 23 multi-wavelength observations of the \ring\ ring from 90 cm (350 MHz) to 0.25 nm (1200 PHz) to investigate the nature of the source (Section \ref{multiwave}). We also compare the magnetic field pseudovectors to the gas kinematics, using previous observations of the HNCO (4-3) transition (Section \ref{shell}) and argue that the magnetic field in \ring\ is being compressed by a driven radial expansion (Section \ref{bfield-dis}). Finally, we explore potential point-source candidates that could be associated with the driver of such an expansion (Section \ref{point-source}).

\subsection{Nature of the M0.8-0.2 Ring}
\label{multiwave}

The circular morphology of \ring\ suggests the nature of the source is some type of bubble, probably either an \hii\ region or a supernova remnant \citep[\eg,][]{Pierce-Price00}. 
\cite{tsuboi15} investigated the kinematics of the source, using SiO and H$^{13}$CO$^+$, to measure the energy needed to produce the expansion velocities observed in their data and argue that a supernova explosion is needed to power the expanding shell. We further explore the nature of \ring\ by undertaking a multi-wavelength approach using the numerous surveys conducted on the CMZ (Table \ref{Images}).

\begin{table*}[bt!]
\caption{\textbf{List of survey datasets that cover the \ring\ ring. The wavelengths shown in bold font correspond to the datasets included in Figure \ref{panel}.}}
\centering
\begin{tabular}{cccccc}
\hline\hline
\multicolumn{2}{c}{\textbf{Wavelength}} & \textbf{Resolution} & \textbf{Telescope} & \textbf{Publication} & \textbf{Notes} \\
\hline
& \textbf{90 cm} & 43\arcsec$\times$24\arcsec & VLA & \cite{LaRosa00} & 350 MHz \\ 
\textbf{Radio} & \textbf{20 cm}\footnote{This dataset, shown in Figure \ref{panel}, has been smoothed to 7\farcs2. } & 4\arcsec & MeerKAT & \cite{heywood22} & 1 GHz  \\ 
& \textbf{6.0 cm} & 9\arcsec$\times$4\arcsec & VLA & \cite{White05} & GPS, C band, 5 GHz \\
& \textbf{1.3 cm} & 26.2\arcsec$\times$17.8\arcsec & ATCA & J. Ott (private communication) & SWAG, K band, 25 GHz \\
\hline
& \textbf{3.0 mm}  & 9\arcsec & GBT & \cite{ginsburg20} & 90 GHz, MUSTANG, MGPS \\
& \textbf{2.0 mm}  & 21\arcsec & IRAM-30m & \cite{arendt19} & GISMO, 150 GHz \\
\textbf{Microwave}  & \textbf{1.1 mm}  & 33\arcsec & Bolocam & \cite{bally10} &  \til220 GHz, (BGPS)\\ 
& \textbf{870 \micron}  & 18.2\arcsec & APEX-12m & \cite{atlasgal} & ATLASGAL, LABOCA  \\
&850 \micron & 13\arcsec & JCMT & \cite{parsons18} & SCUBA2, 347 GHz \\
\hline
& \textbf{500 \micron}  & 35.0\arcsec & Herschel & \cite{molinari10, molinari11,molinari16} & SPIRE, Hi-GAL\\
&450 \micron & 8\arcsec & JCMT & \cite{parsons18} & SCUBA2, 664 GHz \\
& \textbf{350 \micron}  & 24.0\arcsec & Herschel & \cite{molinari10, molinari11,molinari16} & SPIRE, Hi-GAL \\
\textbf{Far IR} & \textbf{250 \micron}  & 18.0\arcsec & Herschel & \cite{molinari10, molinari11,molinari16} & SPIRE Hi-GAL\\
& \textbf{214 \micron} & 18.2\arcsec & SOFIA & \cite{my23} & HAWC+, Band E \\
& \textbf{160 \micron}  & 12.0\arcsec & Herschel & \cite{molinari10, molinari11,molinari16} & PACS, Hi-GAL\\
& \textbf{70 \micron}  & 6.0\arcsec & Herschel & \cite{molinari10, molinari11,molinari16} & PACS, Hi-GAL\\
&70 \micron & 18.0\arcsec & Spitzer & \cite{carey09}& MIPSGAL  \\
\hline
& \textbf{24 \micron} & 6.0\arcsec & Spitzer & \cite{carey09}& MIPSGAL  \\
& \textbf{22 \micron} & 12\arcsec & WISE & \cite{Wright10} & W4 band \\
&21 \micron & 20\arcsec & MSX & \cite{Egan98} & Band E\\
\textbf{Mid IR}  & \textbf{15 \micron} & 6\arcsec & ISO & \cite{Omont03} & ISOGAL, DENIS  \\
&15 \micron & 20\arcsec & MSX & \cite{Egan98} & Band D\\
&12 \micron & 20\arcsec & MSX & \cite{Egan98} & Band C\\
& \textbf{12 \micron}  & 6.5\arcsec & WISE & \cite{Wright10} & W3 band \\
\hline
&8.3 \micron & 20\arcsec & MSX & \cite{Egan98} &  Band A \\
& \textbf{8.0 \micron}  & 1.2\arcsec & Spitzer & \cite{Churchwell09} & GLIMPSE II, IRAC4 \\
& \textbf{7.8 \micron}  & 6\arcsec & ISO & \cite{Omont03}  & ISOGAL, DENIS \\
& \textbf{5.8 \micron} & 1.2\arcsec & Spitzer & \cite{Churchwell09} & GLIMPSE II, IRAC3 \\
&4.6 \micron & 6.4\arcsec & WISE & \cite{Wright10} & W2 band \\
\textbf{Near IR} & \textbf{4.5 \micron} & 1.2\arcsec & Spitzer & \cite{Churchwell09}  & GLIMPSE II, IRAC2 \\
&\textbf{3.6 \micron} & 1.2\arcsec & Spitzer & \cite{Churchwell09}  & GLIMPSE II, IRAC1 \\
& 3.4 \micron & 6.1\arcsec & WISE & \cite{Wright10} & W1 band \\
&2.159 \micron & 2.0\arcsec & 2MASS & \cite{2mass} & 2MASS, Band K \\ 
&1.662 \micron & 2.0\arcsec & 2MASS & \cite{2mass}   & 2MASS, Band H \\ 
&1.235 \micron & 2.0\arcsec & 2MASS & \cite{2mass}   & 2MASS, Band J \\ 
\hline
 & 1.0 nm & 0.5\arcsec & Chandra & \cite{wang21} & 1--4 kEV \\
\textbf{X-ray} & \textbf{0.25 nm} & 0.5\arcsec & Chandra & \cite{wang21} & 4--6 kEV \\
 & 0.12 nm & 0.5\arcsec & Chandra & \cite{wang21} & 6--9 kEV  \\
\hline \hline
\end{tabular}
\label{Images}
\end{table*}

Figure \ref{panel} shows the emission from the ring across 19 different wavelengths, from 90 cm (350 MHz) to 0.25 nm (1200 PHz). The information for the images shown in Figure \ref{panel} is listed in Table \ref{Images}, including the angular resolution (column 2), telescope (column 3), and publication (column 4). We also include additional surveys of the region in Table \ref{Images} that are not shown in Figure \ref{panel}, either due to redundancy with other similar observations (e.g., 870 \micron\ ATLASGAL and 850 \micron\ SCUBA2 images) or because they are non-detections (e.g., 2MASS observations).

The \ring\ ring, shown in Figure \ref{3color}, is detected in the microwave and far-infrared continua between 3 mm and 160 \micron\ (Figure \ref{panel}). Large-scale emission from the \ring\ ring is not observed at wavelengths shorter than 160 \micron, indicating that the emitting dust is relatively cold. Indeed, \cite{Marsh2015, Marsh2017} and \cite{molinari16} measured the dust temperature across the CMZ and found that the dust temperature in \ring\ is \til17 K.

Between 70 \micron\ and 3 mm, there is a depression in the continuum intensity toward the center of the cloud, with a radius of 1.5\arcmin\ (3.5 pc). However, as shown in many of the observations in Figure \ref{panel} (e.g., 2 mm, 250 \micron, and 160 \micron), this depression is not entirely a vacated cavity. Similarly, this is true for the molecular line emission as well; as discussed in Section \ref{shell}, faint molecular emission can be observed toward the center of the \ring\ ring.

At radio wavelengths of 3 mm and 2 mm, there is more extended emission towards the southeast region of the cloud that is not as prominent in the shorter far-IR wavelength datasets (i.e., 214 \micron\ and 160 \micron\ observations). This extended emission is also visible in Figure \ref{3color}, could be associated with thermal free-free emission associated with an \hii\ region. {Future analysis using higher resolution data from ongoing CMZ surveys (e.g., ACES, JACKS) may give insight on the nature of this radio emission.}

\begin{figure}
\centering
\includegraphics[width=0.47\textwidth]{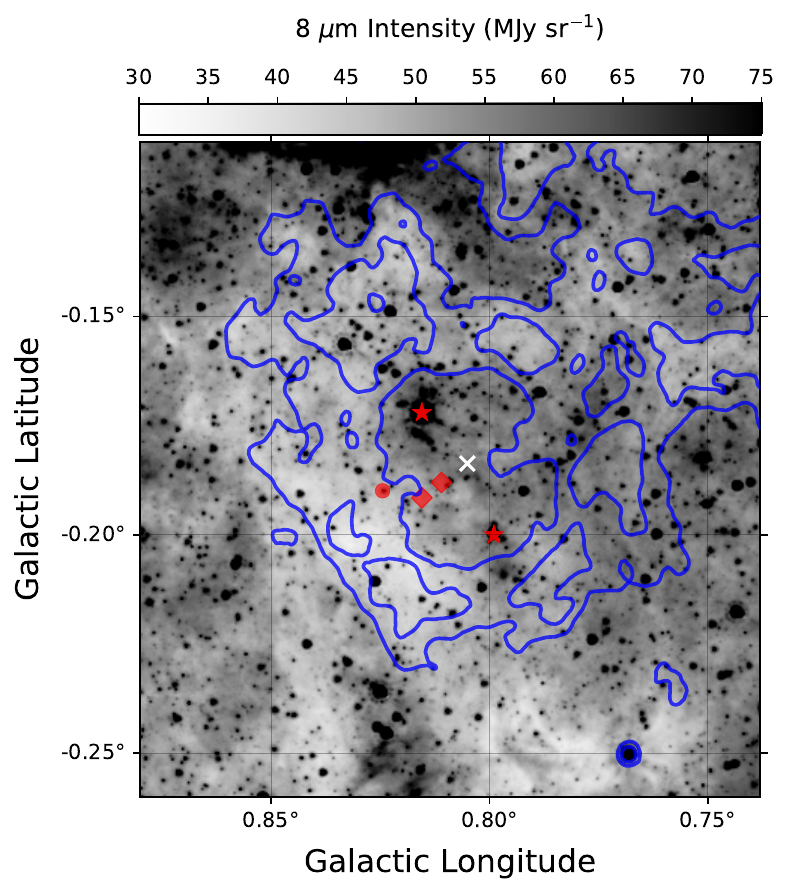}
\caption{Spitzer 8 \micron\ emission \citep{Churchwell09}, scaled (linearly) from 30 to 75 \Mjsr\ and shown in greyscale.
Overlaid are blue contours showing the Herschel+PACS 160 \micron\ emission at 4.0, 4.5, and 5.0 $\times$ Jy/pixel \citep{molinari16}. The locations of several prominent point-sources, discussed in Section \ref{point-source}, are marked.  The two red star markers highlight the two featured IR sources, which are visible in this image. The red circle shows the location of the radio point source. 8 \micron\ emission from the radio point source is also visible in this figure. The two red diamonds show the locations of the two X-ray point sources nearest to the ring center. The white `$\times$' symbol marks the location of the (r=0, $\theta$=0) location used in the polar coordinate system, presented in Figure \ref{unwrap}.}
\label{dark-cloud}
\end{figure}

There is a large clump of IR emission towards the Northeast region of the cloud (labeled as the `stone' in the top left panel of Figure \ref{panel}). This region contains relatively low fractional polarization in the 214 \micron\ emission (Figure \ref{ring}). Relatively low fractional polarization, towards a brighter 214 \micron\ emission region is observed towards other clouds in the FIREPLACE I survey (e.g., Sgr B2, Cloud E/F, Sgr B2NW). The low fractional polarization in these other CMZ sources is argued to be due to: 1) superposition of multiple field directions, 2) grain alignment in the cloud interior may be less efficient, and/or 3) tangling of the field lines below our resolution \citep{my23}. The low fractional polarization in the `stone' could be an indication that similar phenomena are happening here.

There is also a small gap in the far-IR emission around $\theta$=300\degree\ (Figure \ref{ring-polar}). This could be caused by a blow out \citep[a champagne flow; \eg,][]{Deharveng10} or potentially a disruption in the structure. This region also shows relatively higher emission in the 8 \micron\ Spitzer data than other lines-of-sight towards \ring. This relatively higher emission observed in the Spitzer data is also observed in other mid-IR and near-IR panels, covering a wavelength range of \til4--15 \micron.

\begin{figure}[t!]
\centering
\includegraphics[width=0.475\textwidth]{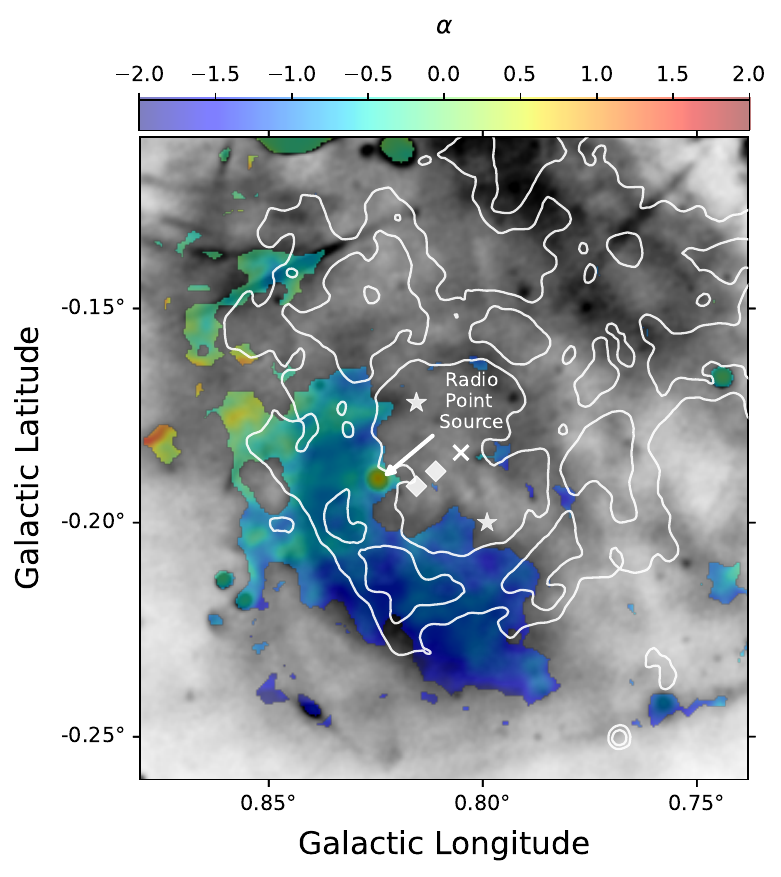}
\caption{MeerKAT 1 GHz radio emission from \cite{heywood22} showing the radio spectral index, $\alpha$, from $-$2 to +2, in colorscale, where $S\propto\nu^{\alpha}$, overlaid on the total intensity: $S$ (Stokes $I$, greyscale). The Stokes $I$ image has been smoothed to an angular resolution of 7\farcs2. The white contours show the Herschel+PACS 160 \micron\ emission at 4.0, 4.5 and 5.0 $\times$ Jy/pixel \citep{molinari16}. The radio point source is annotated in the figure with the white arrow. The locations of several other prominent point-sources, discussed in Section \ref{point-source}, are also marked. The two white star markers highlight the two featured IR sources. The two white diamonds show the locations of the two X-ray point sources nearest to the ring center. The white `$\times$' symbol marks the location of the (r=0, $\theta$=0) location used in the polar coordinate system, presented in Figure \ref{unwrap}. }
\label{SI}
\end{figure}

At wavelengths shorter than 70 \micron\ the ring is observed to be IR dark (bottom two rows in Figure \ref{panel}), in that the emission towards \ring\ is lower than other regions in the field-of-view. The notion that \ring\ is an IR dark cloud is consistent with T$_{\rm dust}\lesssim$20 K reported in \cite{molinari11}. Figure \ref{dark-cloud} shows a comparison of the Spitzer 8 \micron\ emission (greyscale) to the Herschel 160 \micron\ emission (contours). The \ring\ ring, as shown by the contours, generally traces the extent of the dark cloud (greyscale), indicating \ring\ is optically thick at mid- and near-IR wavelengths.

\subsubsection{Radio Emission}
\label{radio}

At radio wavelengths (90 cm -- 1 cm; top row of Figure \ref{panel}), there is very little emission associated with the \ring\ ring. The recent Galactic Center MeerKAT data release \citep{heywood22} shows two diffuse emission regions around the northeastern and southern region of the cloud. These emission regions are located in the same general direction as the extended emission in the millimeter wavelengths (i.e., 3 mm and 2 mm panels), discussed above. We could also be observing these emission regions in the 90 cm panel, however, the resolution of these observations is very course (\til25--45\arcsec).

The 1.28 GHz study by \cite{heywood22} reported the spectral index of the detected radio sources across their observed bandwidth. Figure \ref{SI} shows the spectral index of the ring at radio wavelengths using the dataset from \cite{heywood22}.\footnote{\hyperlink{https://archive-gw-1.kat.ac.za/public/repository/10.48479/fyst-hj47/index.html}{Link to the \cite{heywood22} MeerKAT dataset.}}
As shown, the spectral index for \ring\ is $\alpha$\til$-$0.8 ($S\propto\nu^{\alpha}$), with values ranging from $-$0.4 to $-$1.2, with the southern region of the source containing a steeper spectral index. This relatively steep spectrum in the diffuse emission is consistent with the other radio observations of this source (Figure \ref{panel}). The lowest frequency observations \citep[90 cm, 350 MHz;][]{LaRosa00} show radio emission from \ring\ that is relatively brighter towards the Southern region compared to the Northeastern region. This negative spectral index is consistent with the non-detection of diffuse emission in the 6 cm and 1 cm observations. The negative spectral index at radio wavelengths suggests the emission observed at longer radio wavelengths (90 cm and 20 cm) is likely associated with non-thermal synchrotron radiation. Synchrotron emission is commonly associated with supernova remnants, which produce the relativistic electrons needed to produce the radiation. The detection of synchrotron emission at radio wavelengths, implies that \ring\ is likely to be a supernova remnant,  consistent with the suggestion by \cite{tsuboi15}.

Assuming the emission observed in the MeerKAT 1 GHz radio continuum is synchrotron radiation, as suggested by the negative spectral index, we can calculate the lifetime ($\tau$) of the cosmic ray electrons producing this radio emission. Here we can estimate the lifetime $\tau\equiv E/P$, where $E=m_ec^2\gamma$ is the relativistic electron energy and $P= (4/3)c\sigma_TU_B\gamma^2$, (with $\sigma_T$ the Thomson cross-section {and $U_B$ is the magnetic field energy density}) is the power radiated, of the synchrotron emitting electrons.  We make the simplifying assumption that radiation observed at some frequency $\nu$ is radiated by electrons with Lorentz factor $\gamma=(16\nu m_ec/3eB)^{1/2}$. Since we know $B$ (Section \ref{bfield}), this allows us to estimate $\gamma$. The result is 

\begin{equation}
\tau=1.4\times10^6 \biggl(\frac{B}{100\mu \textrm{G}}\biggl)^{-3/2}  \biggl(\frac{\nu_c}{\textrm{GHz}}\biggl)^{-1/2} \textrm{yr}.
\label{Synch6}
\end{equation}
where $B$ and $\nu_c$ are the magnetic field in units of 100 $\mu$G and observed frequency in units of GHz, respectively. 

Using the average magnetic field strength from the SOFIA/HAWC+ observations, derived in Section \ref{bfield} ($B=B_{\rm POS}=572\pm12$ $\mu$G), and $\nu_c$=1.28 GHz from the \cite{heywood22} MeerKAT observations (see Figures \ref{panel} and \ref{SI}), we calculate a lifetime of 0.9($\pm$0.03)$\times$10$^5$ yr using Equation \ref{Synch6}. This lifetime estimate, derived from the synchrotron emission, is roughly within a factor of 2 of the estimate of \cite{tsuboi15} who argue that the upper limit for the age of \ring\ is 1.8$\times$10$^5$ yr, based on calculations from the gas kinematics. 
The indication that \ring\ is an evolved SNR is consistent with the dust temperature measurement derived by \cite{molinari11}. They estimate a relatively cold dust temperature of $\lesssim$20 K in \ring\ - which is consistent with other similarly-aged SNRs and suggestive of radiative IR cooling \citep[\eg,][]{Chawner20,wang22}.
We note, however, that our lifetime estimate, calculated above, does not include bremsstrahlung losses, which could be significant at the characteristic energies of the radiating electrons \citep[e.g., see Figure 2 of][]{Yoast-Hull13}. 
Therefore, we suggest that our estimate of the lifetime be regarded as an upper limit.

\subsubsection{X-Ray Emission}
\label{xray}

Using archival observations from Chandra \citep{wang21}, we examined the diffuse X-ray emission around the \ring\ ring but find no indications of structure across the 1-9 keV band (4-6 keV is shown in the bottom right panel of Figure \ref{panel}). To put limits on the extended surface brightness of X-ray emission from the ring and its interior, such as might arise from a supernova remnant (SNR), we focused specifically on the integrated 2-4 keV emission. This band was chosen because SNR X-ray emission is relatively soft \citep[e.g.,][]{Vink17}, especially from older remnants, and because foreground Galactic absorption cuts the spectrum off rapidly below 2 keV \citep[e.g.,][]{Baganoff+03}. For this analysis we use the three regions defined in Section \ref{res}: (I) a region interior to the ring, (II) an annular region matching the “band” structure of the ring identified in the mid-IR images, and (III) an exterior annular region extending $\sim 1$ arcmin beyond the ring. Integrating the 2-4 X-ray fluxes in these regions after removing all significant point sources, and then converting to surface brightness, we find for (I) the inner region $(4.37 \pm 0.06) \times 10^{28}$ erg s$^{-1}$ arcsec$^{-2}$, for (II) the ring $(4.6 \pm 0.1) \times 10^{28}$ erg s$^{-1}$ arcsec$^{-2}$, and for (III) the outer region $(4.8 \pm 0.6) \times 10^{28}$ erg s$^{-1}$ arcsec$^{-2}$. We therefore find no excess surface brightness in the region corresponding to \ring\ or its interior beyond the local ambient X-ray background, which is comprised of a population of X-ray point sources below the detection threshold, the cosmic X-ray background, and any extended hot plasma emission that may be present in this region. The lack of excess X-ray emission within and interior to the \ring\ ring suggests that if it is an SNR, it must be old enough to no longer have appreciable X-ray emission \citep[\eg,][]{Ou18}. The suggestion that \ring\ is an old SNR is consistent with the previous calculations of the synchrotron radio emission, and with the age estimates calculated in \cite{tsuboi15}, discussed previously in Section \ref{radio}. 

\begin{figure*}[p!]
\centering
\includegraphics[width=1.0\textwidth]{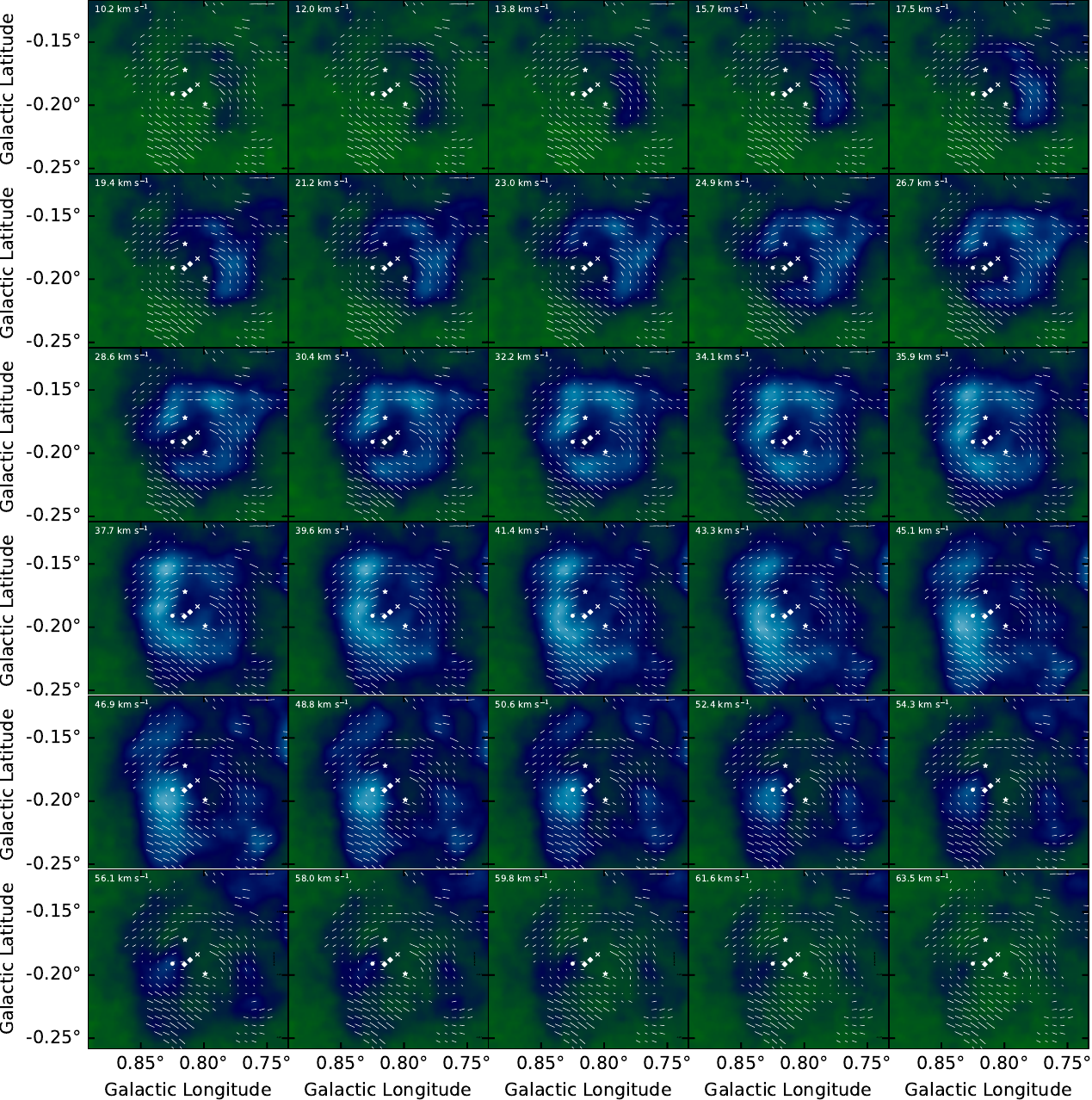}
\caption{
Channel maps of the HNCO (4$-$3) emission from the \cite{jones12} 3 mm Mopra survey, covering a velocity range of 10.2--63.5 \kms. The spatial resolution at the frequency of the HNCO line is 36\arcsec. The spectral resolution of the HNCO (4$-$3) line is 1.84 \kms. The velocity of each channel is annotated in the top left corner of each panel. Overlaid on the HNCO background are the magnetic field pseudovectors shown in Figure \ref{ring}, using Nyquist sampling at the HNCO beamsize, and only the 214 \micron\ pseudovectors for which the intensity is above 5,000 \Mjsr. The locations of several prominent point-sources, discussed in Section \ref{point-source}, are marked. The two white star markers highlight the two featured IR sources. The white circle shows the location of the radio point source. The two white diamonds show the locations of the two X-ray point sources nearest to the ring center. The white `$\times$' symbol marks the location of the (r=0, $\theta$=0) location used in the polar coordinate system, presented in Figure \ref{unwrap}. }
\label{channels}
\end{figure*}

\subsection{Comparison of the magnetic field structure to the gas kinematics}
\label{shell}

The ring-like morphology of the \ring\ cloud is observed in molecular emission as well. Relatively bright SiO (2--1) emission in the ring was detected by \citet[][identified as Shell 5 in their work]{tsuboi15}. The SiO molecule is a known tracer of shocked gas and has been previously detected toward other supernova remnants \citep[\eg,][]{Matsuura17, Abellan17, Cosentino22}, indicating the SiO detection in \cite{tsuboi15} may also have originated from similar phenomena. Similar to SiO, as a known shock tracer, HNCO is also known to trace low-velocity shocks \citep{Yu18}. The \ring\ ring is also {known} to have relatively elevated HNCO emission compared to other clouds in the CMZ \citep{mills17}, indicating the cloud could have elevated shock activity relative to other CMZ clouds. Most of this elevated emission in the HNCO emission, relative to the HCN emission, is located at smaller radii \citep[see Figure 2 in][]{mills17}, indicating that the shocked material is predominantly at the smaller radii. This is consistent with the hypothesis that the ring could be the result of an expanding shock front that is propagating radially outward and is sweeping up material. We further explore the HNCO emission, which is used to derive the magnetic field strength in Section \ref{bfield}, to investigate the magnetic field structure in relation to the gas kinematics.

Figure \ref{channels} shows {channel maps of the} HNCO (J=4$-$3) emission from the 3 mm Mopra survey \citep{jones12, mills17}, overlaid with the magnetic field pseudovectors from the FIREPLACE survey \citep{my23}. To reduce the number of pseudovectors in the image, and allow for a better visual inspection of the data, we have reduced the number of vectors plotted in the image by a factor of 2.5. This approximately corresponds to one pseudovector for each HAWC+ beam at 214 \micron. These channel maps, identified by the velocity in the top left corner, show the clear central depression in the cloud at all velocities represented in the cloud. Furthermore, this structure is coherent in velocity space, indicating this morphology is associated with a unified structure, rather than a line-of-sight arrangement of distinct velocity components.
HNCO emission is brightest towards the southeast region of the cloud, where we also observe some of the brightest emission in the 214 \micron\ continuum.

The kinematics shown in Figure \ref{channels} also provide a consistency check on the $\sigma_v$ radial and azimuthal profiles shown in Figure \ref{fig:disp_analsis_radii}c, which are based on this dataset for the DCF analysis ($\sigma_v$ parameter in Equation \ref{eq:bfield eq_1}). The radial profile shows a generally decreasing velocity dispersion at larger radii. The azimuthal velocity dispersion profile, however, shows two main peaks at \til100\degree\ and \til270\degree\ (corresponding to the East and West regions of the cloud, respectively). This can be discerned from Figure \ref{channels}, which shows excess emission towards the western side of the cloud at lower velocities and an excess emission towards the eastern side of the cloud at higher velocities, resulting in a higher velocity dispersion at these azimuthal angles, as observed in Figure \ref{fig:disp_analsis_radii}c.

The magnetic field pseudovectors appear to closely follow the morphology of the HNCO emission and clearly highlight the line-of-sight velocity components contributing to the integrated polarization. The morphology of the HNCO emission shows a gap in the ring-like structure near the radio point source (marked by the white circle) at velocities between 25--35 \kms. At velocities of 35--40 \kms\ the distribution of molecular emission appears to arc around the point source. At the higher velocities in the ring (i.e., 45--55 \kms, second row from the bottom in Figure \ref{channels}) the HNCO emission is compact and associated closely with the position of the point source. We discuss the possible interaction between \ring\ and nearby point sources in Section \ref{point-source}.

\subsection{Compression of the Magnetic Field in \ring}
\label{bfield-dis}

Flux freezing occurs when the conductivity in the material is sufficiently high that the diffusion time is larger than the dynamical time of the system \citep[\ie, magnetic {Reynold's number:} $R_M\gg 1$; \eg,][]{Mestel1966}. As the source material is displaced through an external force (e.g., gravity) the magnetic field lines are dragged along with the displaced material. This frozen-in flux phenomenon is commonly observed in the ISM and could favor the formation and the internal pressure support of molecular clouds \citep[see][for a recent review on magnetic fields in molecular clouds]{Hennebelle19}.

Feedback effects from massive stars can compress uniform fields, resulting in an enhanced field strength and bubble-like morphologies \citep[\eg,][]{Vrba87, milne90}. \cite{Vrba87} showed that as a magnetic field is compressed, presumably through a shock encounter, the original orientation of the field is shifted from the original $B_0$ direction toward an orientation that is oriented perpendicular to the radial direction (see the schematic in their Figure 6). This is due to the perpendicular field component being continuous (i.e., $\nabla \textbf{B} = 0$), whereas the parallel field component is amplified by the shock compression \citep{Vrba87}. \cite{milne90} also showed a schematic of shock compression of a magnetic field near a SNR in their Figure 1.  In their figure, they show the propagation of the shock front stretching and compressing the magnetic field lines, as the ISM itself is compressed, and highlight regions of X-ray emission located within the SNR.

A ring feature similar to M0.8–0.2 was investigated by \cite{chen17}, who observed a curved magnetic field morphology toward the N4 bubble, using near IR polarimetry. The N4 bubble is a shell structure projected as a ring on the POS.  It encloses the \hii\ region G11.898+0.747 identified by \cite{Lockman89}. Figure 5 in \cite{chen17} shows the N4 bubble, plotted in polar coordinates, similar to our Figure \ref{unwrap}, which shows an analogous distribution of the magnetic field orientations, aligned largely perpendicular to the radial direction. They measure the field outside of the shell and argue for an enhancement in the field strength due to compression.  Similarly, \cite{bracco20} saw an enhancement in the field strength in the Corona Australis molecular cloud that is twice that of the surrounding material. They used Planck polarimetry data \citep{PlanckCollaboration2011} and \hi\ data from GASS \citep{McClure-Griffiths09} to investigate the magnetic field strength at an intersection of two \hi\ shells in the cloud and argue that field compression is causing this enhancement.

We observe a similar field enhancement in the \ring\ ring. The magnetic field strength in the ring increases and then decreases as a function of radius (see Figure \ref{fig:disp_analsis_radii}a, {\it Left}). The field strength at the inner radius, Region I ($<$3 pc; \til220-750 $\mu$G), is lower than the field strength in Region II (\til4--5 pc; \til300-1000 $\mu$G), which is higher than the strength at the largest radii (\til8--9 pc; \til140-500 $\mu$G). The observed field strengths at these largest radii are comparable to magnetic field estimates for other clouds in the CMZ \citep[\eg,][]{Ferriere09}. Assuming the magnetic field strength at the largest radii is similar to the ambient CMZ field strength, we measure a factor of \til2 increase in the field strength due to this compression of the magnetic field in region II (e.g., see Figure \ref{fig:disp_analsis_radii}). This doubling of the field strength due to the compression is similar to the findings by \cite{bracco20}, who also saw a factor of two increase in their magnetic field strength estimates.

The structure of the magnetic field in polar coordinates (Figure \ref{unwrap}) also supports this idea of field compression. The field shows a well organized structure at short radii (Regions I and II), and is less organized at larger radii (Region III, $\gtrsim$7 pc). In regions where the ring-like structure is most prominent (Region II, \til3--7 pc), the field lines are generally perpendicular to the radial direction, indicating that the field lines have a ring-like structure. The presence of a ring-like structure in the magnetic field, combined with the ring-like structure in the dust morphology, is consistent with our scenario in which the magnetic field is being swept up and compressed by the gas and dust in \ring.

We can further compare the orientation of the magnetic field in polar coordinates to the schematic presented in \citet[][their Figure 6]{Vrba87}. For an expanding shell, this shock front would be propagating outward in the radial direction from a central location. In general we find this to be true for magnetic field vectors associated with Region II. There are, however, lower-density regions in this radial range where we observe a deviation from this orientation (\eg, $\theta\simeq$ 0\degree, 70\degree, 180\degree, and 310\degree; Figure \ref{unwrap}). These regions, where we observe the field to deviate from being flat, could represent regions where the field is being influence by the ambient field or by expansion of the shell past a relatively dense, pre-existing gas clump.

The inner radii of the ring, Region I ($<$3 pc), show a higher magnetic field compared to the ambient CMZ field strength, but not as high as Region II. Despite the low column density in Region I ($N(H_2)$ panel in Figure \ref{fig:disp_analsis_radii}b, left), we observe a factor of 1.5 increase in the field strength compared to the ambient field at larger radii (Region III, \til8--9 pc). This inner region of \ring\ also shows relatively high fractional polarization values, up to 20\% (see Figure \ref{p vs. r}), suggesting any dust in this inner region is well aligned with the magnetic field. \cite{Akshaya23} observed highly polarized 214 \micron\ emission in regions with low column density, high temperature and low magnetic field tangling.

\subsection{Potential Sources Powering the \ring\ Ring}
\label{point-source}

There are several point-sources that could be associated with the mechanism producing the expansion of the \ring\ cloud. Observations of the region at radio wavelengths (1--90 GHz) reveal a relatively bright point source located near the inner eastern edge of the shell (\ra=17\h48\m18\fs7, \dec=-28\degree19\arcmin48\arcsec; $l$=0\fdg825, $b$=-0\fdg191; see Figures \ref{3color} and \ref{panel}--\ref{channels}). This point source is clearly detected in the MeerKAT, VLA, ATCA, and GBT datasets (see Figure \ref{panel}), and it is also detected at mid- and far-infrared wavelengths, 8--70 \micron. We note that the point source could be emitting radiation at wavelengths shorter than 8 \micron, however, the source confusion in the Spitzer datasets (3.6--5.8 \micron) and the course sensitivity in the ISOGAL dataset make identifying the radio point source challenging. The emission from the radio point source is not detected at X-ray wavelengths \citep[see Chandra panel in Figure \ref{panel};][]{wang21}. We also note the presence of roughly a dozen other compact radio sources in the vicinity of \ring\ that are observed in the MeerKAT dataset (Figure \ref{panel}). However, due to the relative faintness of these sources, we will not be discussing them in this paper.

The MeerKAT 1 GHz data show the featured point source has a rising spectral index of +0.7, suggesting it is thermal in nature \citep[][see Figure \ref{SI}]{heywood22}. This observed rising spectral index in the MeerKAT data is consistent with the 3 mm detection in the GBT observations. The spectral index of the point source suggests that the radio emission is likely associated with free-free (bremsstrahlung) emission. 
The detection of this source at mid-IR wavelengths (8--70 \micron) suggests there is warm dust that is also associated with the thermal radio point-source (see Figure \ref{dark-cloud}).

This radio point-source is located near one of the convergence locations in the magnetic field lines (see Section \ref{pol-section}, and Figures \ref{ring}--\ref{unwrap}). This location also shows relatively low fractional polarization in the 214 \micron\ emission (see Figure \ref{ring}), which could be indicative of an interaction between the point source and the \ring\ ring.

The diffuse radio emission also observed in the 1 GHz MeerKAT observations (Figure \ref{panel}) is predominately located near, in projection, the radio point-source. This could be an indication that this radio point-source is producing the 1 GHz emission observed here. Additionally, the extended 90 GHz GBT emission (shown in green in Figure \ref{3color}) is also predominantly located towards this region of \ring, and is not observed at far-IR wavelengths, which could be an indication that this emission is being influenced by the radio point-source. This superposition of the radio point source and the diffuse 1 GHz and 90 GHz emission hints that this point source is spatially co-located within \ring\ instead of being just a projection effect. 

Furthermore, the morphology in the HNCO channel maps, shown in Figure \ref{channels}, supports the argument that the radio point source is physically interacting with the \ring\ ring. As discussed in Section \ref{shell}, the morphology of the HNCO emission shows a gap in the ring-like structure at the location of the radio point source around velocities of 25--35 \kms\ and arcs around the point source at velocities of 35--40 \kms. At the highest velocities of \ring, 45--55 \kms, the emission is compact and co-located with the radio point source.

Additionally, there are two IR point-sources that are located, in projection, near the center of the ring (see Figure \ref{3color}, and star markers in Figures \ref{dark-cloud}--\ref{channels}). These two sources are located at \ra=17\h48\m13\fs09,  \dec=-28\degree19\arcmin42\farcs2 ($l$=0\fdg815, $b$=-0\fdg172) and \ra=17\h48\m17\fs31, \dec=-28\degree21\arcmin24\farcs4 ($l$=0\fdg799, $b$=-0\fdg200). The two IR point sources are located at the center of each of these depressions. The projected location of these two IR point sources makes them candidate sources for the expansion of the \ring\ ring. We note, however, that these point sources could easily be a chance alignment due to the low emission towards the center of the ring.  

There are also two X-ray point sources that are located, in projection, within the interior of the \ring\ ring (see diamond markers in Figures \ref{dark-cloud}--\ref{channels}). These X-ray point sources are located at 
$l$=0\fdg816, $b$=-0\fdg192 and 
$l$=0\fdg811, $b$=-0\fdg188. 
These X-ray sources could be clumps associated with the interior of \ring, assuming it is a SNR, as indicated in \cite{milne90} (see their Figure 1). Follow up analysis of these X-ray, IR, and radio point sources is necessary to determine whether they are interacting with or produced by, the \ring\ ring. Distance determination of each is critical for identifying them as GC sources. This follow up analysis would also be beneficial to understanding the physics of these sources to determine if they could produce the energy needed to drive the expansion, as derived in \cite{tsuboi15}.

\section{Summary}
\label{conclusion}

We present the detection of a polarized dust ring (\ring) in the Galactic Center as part of the Far-Infrared Polarimetric Large Area CMZ Emission (FIREPLACE) DR1 legacy survey \citep{my23}. The survey was taken with the 214 \micron\ band of the HAWC+ instrument (19\farcs6 angular resolution) aboard SOFIA. This region contains a circular magnetic field that generally traces the ring-like structure of the cloud (Figure \ref{ring}). 

We illustrate the azimuthal nature of the magnetic field by plotting the vectors in a polar coordinate system (see Figure \ref{unwrap} and Section \ref{pol-section}). Using the DCF method, we measure the magnetic field strength as a function of radius and a function of azimuthal angle and observe that the field strength generally decreases as a function of larger radii (see Section \ref{bfield}). However, the magnetic field strength locally increases at $r$$\sim$4--5 pc (see Figure \ref{fig:disp_analsis_radii}). These locations correspond to radii where the column density also increases, suggesting the magnetic fields are compressed. 

We further compare our results to a multi-wavelength analysis of the source by utilizing the many surveys of the CMZ (see Table \ref{Images} for this listing of surveys, Figure \ref{panel} for a visual subset of these datasets, and Section \ref{multiwave} for the corresponding discussion). We investigate the identify and nature of the driving source using this multi-wavelength approach, combined with kinematic analysis presented in \cite{tsuboi15}, who argue the \ring\ ring is a supernova remnant. The multi-wavelength analysis, and the spectral index results from \cite{heywood22}, which shows a falling spectral index (Figure \ref{SI}), support this argument that \ring\ is a supernova remnant. 

We also compare the 214 \micron\ magnetic field pseudovectors to the HNCO emission, a known shock tracer, illustrated as channel maps from 10.2--63.5 \kms\ (Figure \ref{channels}). The magnetic field generally follows the gas kinematics and morphology, suggesting the circular-like nature of the magnetic field is produced by the expansion of the supernova remnant. Furthermore, this expansion of \ring\ may also be compressing the magnetic fields, causing the enhanced field strengths observed in the cloud (Sections \ref{shell} and \ref{bfield-dis}). 

Lastly, we explore potential sources that could be producing this expansion (Section \ref{point-source}). In particular, the featured radio point source 
is a strong candidate for a possible interaction with \ring. The HNCO emission appear to follow an arc geometry around the location of the point source at lower velocities (v$<$45 \kms) and show a brightening at higher velocities (v$>$45 \kms), suggesting a possible interaction with the molecular gas (Figure \ref{channels}). The magnetic fields at this location also show lower factional polarization values and appear to curve around the location of the point source, which could indicate an interaction between the point source and magnetic fields. 
While the radio point source suggests a possible interaction with \ring\ there are several other potential candidates that cannot be ruled out (Section \ref{point-source}). Follow-up observations of the radio point source, and other nearby point sources, is needed to identify the source(s) or phenomena that have led to the expansion of \ring.

\section{Acknowledgements}

The authors would like to thank Dr. Daniel Wang (University of Massachusetts) for use of his Chandra data, which is shown in Figure \ref{panel}. The authors would also like to thank Dr. Juergen Ott (NRAO) for use of his preliminary SWAG data which is also shown in Figure \ref{panel}. The authors would also like to thank Dr. Ashley Barnes (ESO), Dr. Katharina Immer (ESO), and Mairi Nonhebel (ESO) for their helpful discussion on this source. We would like to thank Dr. Simon Coude, Dr. Sachin Shenoy, Dr. Peter Ashton and the rest of the SOFIA team for their help with the original data reduction. We would also like to the thank the anonymous referee for their helpful review of this work.

\subsection{Telescope \& Survey Acknowledgements}

The main focus of this work (polarized dust emission and magnetic field strength in \ring; Section \ref{res}) is based on observations made with the NASA/DLR Stratospheric Observatory for Infrared Astronomy (SOFIA). SOFIA is jointly operated by the Universities Space Research Association, Inc. (USRA), under NASA contract NNA17BF53C, and the Deutsches SOFIA Institut (DSI) under DLR contract 50 OK 2002 to the University of Stuttgart. Financial support for this work was provided by NASA through awards \#09-0054 and \#09-0537 issued by USRA to Villanova University.

Since we are showing a multi-wavelength dataset in Figure \ref{panel}, which encompasses many surveys, we have included acknowledgements for these other telescopes and/or surveys, in alphabetical order, below:

\begin{itemize}

\item \textbf{ATCA:} 
The Australia Telescope Compact Array is part of the Australia Telescope National Facility (\hyperlink{https://ror.org/05qajvd42}{https://ror.org/05qajvd42}) which is funded by the Australian Government for operation as a National Facility managed by CSIRO. We acknowledge the Gomeroi people as the Traditional Owners of the Observatory site.

\item \textbf{APEX/ATLASGAL:} This publication is based on data acquired with the Atacama Pathfinder Experiment (APEX) under the ATLASGAL large programme \citep{atlasgal}. APEX is a collaboration between the Max-Planck-Institut fur Radioastronomie, the European Southern Observatory, and the Onsala Space Observatory.

\item \textbf{Bolocam/BGPS:} The Bolocam Galactic Plane Survey (BGPS) project is supported in part by the National Science Foundation through NSF grant AST-0708403.

\item \textbf{Chandra:} The scientific results reported in this article are based in part or to a significant degree on observations made by the Chandra X-ray Observatory for articles by the PI team, data obtained from the Chandra Data Archive for articles based on archival data, observations made by the Chandra X-ray Observatory and published previously in cited articles for articles based on published results.

\item \textbf{IRAM-30m:}
IRAM is supported by INSU/ CNRS (France), MPG (Germany) and IGN (Spain).

\item \textbf{ISO:} 
This work is based on observations with ISO, an ESA project with instruments funded by ESA Member States (especially the PI countries: France, Germany, the Netherlands and the United Kingdom) and with the participation of ISAS and NASA.

\item \textbf{GBT/MUSTANG2:} The Green Bank Observatory is a facility of the National Science Foundation operated under cooperative agreement by Associated Universities, Inc. The MUSTANG2 instrument is supported by the NSF award number 1615604 and by the Mt. Cuba Astronomical Foundation. 

\item \textbf{Herschel:} 
Herschel is an ESA space observatory with science instruments provided by European-led Principal Investigator consortia and with important participation from NASA.

\item \textbf{LABOCA:} 
The Large APEX BOlometer CAmera LABOCA”, G. Siringo et al., Astronomy and Astrophysics, Volume 497, Issue 3, 2009, pp.945-962 [2009A\&A…497..945S].

\item \textbf{MeerKAT:} The MeerKAT telescope is operated by the South African Radio Astronomy Observatory, which is a facility
of the National Research Foundation, an agency of the Department of Science and Innovation.

\item \textbf{Mopra:} 
The Mopra telescope is part of the Australia Telescope National Facility (\hyperlink{https://ror.org/05qajvd42}{https://ror.org/05qajvd42}) which is funded by the Australian Government for operation as a National Facility managed by CSIRO. We acknowledge the Gamilaroi people as the Traditional Owners of the Observatory site. The University of New South Wales Digital Filter Bank used for the observations with the Mopra Telescope was provided with support from the Australian Research Council.

\item \textbf{SOFIA/HAWC+:}
``HAWC+, the Far-Infrared Camera and Polarimeter for SOFIA"
\citep{Harper2018}, JAI, 7, 1840008-1025. DOI: 10.1142/S2251171718400081 ; ADS Bibliographic Code: 2018JAI.....740008H. See SOFIA citation above.  

\item \textbf{Spitzer:} This work is based in part on observations made with the Spitzer Space Telescope, which is operated by the Jet Propulsion Laboratory, California Institute of Technology under a contract with NASA.

\item \textbf{VLA:} The Karl G. Jansky Very Large Array (VLA) is a telescope operated by the National Radio Astronomy Observatory. The National Radio Astronomy Observatory is a facility of the National Science Foundation operated under cooperative agreement by Associated Universities, Inc.

\item \textbf{WISE:} 
This publication makes use of data products from the Wide-field Infrared Survey Explorer (WISE), which is a joint project of the University of California, Los Angeles, and the Jet Propulsion Laboratory/California Institute of Technology, funded by the National Aeronautics and Space Administration.

\end{itemize}

\facility{SOFIA, GBT, MeerKAT, Spitzer, Mopra, Herschel, Chandra}

\software{ \texttt{python, Ipython} \citep{Perez2007}, \texttt{numpy} \citep{vanderWalt2011}, \texttt{scipy} \citep{Jones2001} \texttt{matplotlib} \citep{Hunter2007}, \texttt{emcee} \citep{Foreman-Mackey2013}, \texttt{corner} \citep{Foreman-Mackey2016}, \texttt{astropy} \citep{Astropy2013,Astropy2018}, \texttt{Uncertainties} \citep{LebigotUnc}, \texttt{casa} \citep{Casa, 2011ascl.soft07013I}, }

\bibliographystyle{aasjournal}
\bibliography{ring.bib}

\begin{thebibliography}{}
\expandafter\ifx\csname natexlab\endcsname\relax\def\natexlab#1{#1}\fi
\providecommand{\url}[1]{\href{#1}{#1}}
\providecommand{\dodoi}[1]{doi:~\href{http://doi.org/#1}{\nolinkurl{#1}}}
\providecommand{\doeprint}[1]{\href{http://ascl.net/#1}{\nolinkurl{http://ascl.net/#1}}}
\providecommand{\doarXiv}[1]{\href{https://arxiv.org/abs/#1}{\nolinkurl{https://arxiv.org/abs/#1}}}

\bibitem[{{Abell{\'a}n} {et~al.}(2017){Abell{\'a}n}, {Indebetouw}, {Marcaide},
  {Gabler}, {Fransson}, {Spyromilio}, {Burrows}, {Chevalier}, {Cigan},
  {Gaensler}, {Gomez}, {Janka}, {Kirshner}, {Larsson}, {Lundqvist}, {Matsuura},
  {McCray}, {Ng}, {Park}, {Roche}, {Staveley-Smith}, {van Loon}, {Wheeler}, \&
  {Woosley}}]{Abellan17}
{Abell{\'a}n}, F.~J., {Indebetouw}, R., {Marcaide}, J.~M., {et~al.} 2017,
  \apjl, 842, L24, \dodoi{10.3847/2041-8213/aa784c}

\bibitem[{{Akshaya} \& {Hoang}(2023)}]{Akshaya23}
{Akshaya}, M.~S., \& {Hoang}, T. 2023, \mnras, 522, 4196,
  \dodoi{10.1093/mnras/stad1246}

\bibitem[{{Andersson} {et~al.}(2015){Andersson}, {Lazarian}, \&
  {Vaillancourt}}]{Andersson15}
{Andersson}, B.~G., {Lazarian}, A., \& {Vaillancourt}, J.~E. 2015, \araa, 53,
  501, \dodoi{10.1146/annurev-astro-082214-122414}

\bibitem[{{Arendt} {et~al.}(2019){Arendt}, {Staguhn}, {Dwek}, {Morris},
  {Yusef-Zadeh}, {Benford}, {Kov{\'a}cs}, \& {Gonzalez-Quiles}}]{arendt19}
{Arendt}, R.~G., {Staguhn}, J., {Dwek}, E., {et~al.} 2019, \apj, 885, 71,
  \dodoi{10.3847/1538-4357/ab451c}

\bibitem[{{Astropy Collaboration} {et~al.}(2013){Astropy Collaboration},
  {Robitaille}, {Tollerud}, {Greenfield}, {Droettboom}, {Bray}, {Aldcroft},
  {Davis}, {Ginsburg}, {Price-Whelan}, {Kerzendorf}, {Conley}, {Crighton},
  {Barbary}, {Muna}, {Ferguson}, {Grollier}, {Parikh}, {Nair}, {Unther},
  {Deil}, {Woillez}, {Conseil}, {Kramer}, {Turner}, {Singer}, {Fox}, {Weaver},
  {Zabalza}, {Edwards}, {Azalee Bostroem}, {Burke}, {Casey}, {Crawford},
  {Dencheva}, {Ely}, {Jenness}, {Labrie}, {Lim}, {Pierfederici}, {Pontzen},
  {Ptak}, {Refsdal}, {Servillat}, \& {Streicher}}]{Astropy2013}
{Astropy Collaboration}, {Robitaille}, T.~P., {Tollerud}, E.~J., {et~al.} 2013,
  \aap, 558, A33, \dodoi{10.1051/0004-6361/201322068}

\bibitem[{{Astropy Collaboration} {et~al.}(2018){Astropy Collaboration},
  {Price-Whelan}, {Sip{\H{o}}cz}, {G{\"u}nther}, {Lim}, {Crawford}, {Conseil},
  {Shupe}, {Craig}, {Dencheva}, {Ginsburg}, {VanderPlas}, {Bradley},
  {P{\'e}rez-Su{\'a}rez}, {de Val-Borro}, {Aldcroft}, {Cruz}, {Robitaille},
  {Tollerud}, {Ardelean}, {Babej}, {Bach}, {Bachetti}, {Bakanov}, {Bamford},
  {Barentsen}, {Barmby}, {Baumbach}, {Berry}, {Biscani}, {Boquien}, {Bostroem},
  {Bouma}, {Brammer}, {Bray}, {Breytenbach}, {Buddelmeijer}, {Burke},
  {Calderone}, {Cano Rodr{\'\i}guez}, {Cara}, {Cardoso}, {Cheedella}, {Copin},
  {Corrales}, {Crichton}, {D'Avella}, {Deil}, {Depagne}, {Dietrich}, {Donath},
  {Droettboom}, {Earl}, {Erben}, {Fabbro}, {Ferreira}, {Finethy}, {Fox},
  {Garrison}, {Gibbons}, {Goldstein}, {Gommers}, {Greco}, {Greenfield},
  {Groener}, {Grollier}, {Hagen}, {Hirst}, {Homeier}, {Horton}, {Hosseinzadeh},
  {Hu}, {Hunkeler}, {Ivezi{\'c}}, {Jain}, {Jenness}, {Kanarek}, {Kendrew},
  {Kern}, {Kerzendorf}, {Khvalko}, {King}, {Kirkby}, {Kulkarni}, {Kumar},
  {Lee}, {Lenz}, {Littlefair}, {Ma}, {Macleod}, {Mastropietro}, {McCully},
  {Montagnac}, {Morris}, {Mueller}, {Mumford}, {Muna}, {Murphy}, {Nelson},
  {Nguyen}, {Ninan}, {N{\"o}the}, {Ogaz}, {Oh}, {Parejko}, {Parley}, {Pascual},
  {Patil}, {Patil}, {Plunkett}, {Prochaska}, {Rastogi}, {Reddy Janga},
  {Sabater}, {Sakurikar}, {Seifert}, {Sherbert}, {Sherwood-Taylor}, {Shih},
  {Sick}, {Silbiger}, {Singanamalla}, {Singer}, {Sladen}, {Sooley},
  {Sornarajah}, {Streicher}, {Teuben}, {Thomas}, {Tremblay}, {Turner},
  {Terr{\'o}n}, {van Kerkwijk}, {de la Vega}, {Watkins}, {Weaver}, {Whitmore},
  {Woillez}, {Zabalza}, \& {Astropy Contributors}}]{Astropy2018}
{Astropy Collaboration}, {Price-Whelan}, A.~M., {Sip{\H{o}}cz}, B.~M., {et~al.}
  2018, \aj, 156, 123, \dodoi{10.3847/1538-3881/aabc4f}

\bibitem[{{Baganoff} {et~al.}(2003){Baganoff}, {Maeda}, {Morris}, {Bautz},
  {Brandt}, {Cui}, {Doty}, {Feigelson}, {Garmire}, {Pravdo}, {Ricker}, \&
  {Townsley}}]{Baganoff+03}
{Baganoff}, F.~K., {Maeda}, Y., {Morris}, M., {et~al.} 2003, \apj, 591, 891

\bibitem[{{Bally} {et~al.}(1987){Bally}, {Stark}, {Wilson}, \&
  {Henkel}}]{bally87}
{Bally}, J., {Stark}, A.~A., {Wilson}, R.~W., \& {Henkel}, C. 1987, \apjs, 65,
  13, \dodoi{10.1086/191217}

\bibitem[{{Bally} {et~al.}(1988){Bally}, {Stark}, {Wilson}, \&
  {Henkel}}]{bally88}
---. 1988, \apj, 324, 223, \dodoi{10.1086/165891}

\bibitem[{{Bally} {et~al.}(2010){Bally}, {Aguirre}, {Battersby}, {Bradley},
  {Cyganowski}, {Dowell}, {Drosback}, {Dunham}, {Evans}, {Ginsburg}, {Glenn},
  {Harvey}, {Mills}, {Merello}, {Rosolowsky}, {Schlingman}, {Shirley},
  {Stringfellow}, {Walawender}, \& {Williams}}]{bally10}
{Bally}, J., {Aguirre}, J., {Battersby}, C., {et~al.} 2010, \apj, 721, 137,
  \dodoi{10.1088/0004-637X/721/1/137}

\bibitem[{{Bracco} {et~al.}(2020){Bracco}, {Bresnahan}, {Palmeirim},
  {Arzoumanian}, {Andr{\'e}}, {Ward-Thompson}, \& {Marchal}}]{bracco20}
{Bracco}, A., {Bresnahan}, D., {Palmeirim}, P., {et~al.} 2020, \aap, 644, A5,
  \dodoi{10.1051/0004-6361/202039282}

\bibitem[{{Butterfield} {et~al.}(2018){Butterfield}, {Lang}, {Morris}, {Mills},
  \& {Ott}}]{my18}
{Butterfield}, N., {Lang}, C.~C., {Morris}, M., {Mills}, E. A.~C., \& {Ott}, J.
  2018, \apj, 852, 11, \dodoi{10.3847/1538-4357/aa886e}

\bibitem[{{Butterfield} {et~al.}(2023){Butterfield}, {Chuss}, {Guerra},
  {Morris}, {Pare}, {Wollack}, {Dowell}, {Hankins}, {Siah}, {Staguhn}, \&
  {Zweibel}}]{my23}
{Butterfield}, N.~O., {Chuss}, D.~T., {Guerra}, J.~A., {et~al.} 2023, arXiv
  e-prints, arXiv:2306.01681, \dodoi{10.48550/arXiv.2306.01681}

\bibitem[{{Carey} {et~al.}(2009){Carey}, {Noriega-Crespo}, {Mizuno}, {Shenoy},
  {Paladini}, {Kraemer}, {Price}, {Flagey}, {Ryan}, {Ingalls}, {Kuchar},
  {Pinheiro Gon{\c{c}}alves}, {Indebetouw}, {Billot}, {Marleau}, {Padgett},
  {Rebull}, {Bressert}, {Ali}, {Molinari}, {Martin}, {Berriman}, {Boulanger},
  {Latter}, {Miville-Deschenes}, {Shipman}, \& {Testi}}]{carey09}
{Carey}, S.~J., {Noriega-Crespo}, A., {Mizuno}, D.~R., {et~al.} 2009, \pasp,
  121, 76, \dodoi{10.1086/596581}

\bibitem[{{Chandrasekhar} \& {Fermi}(1953)}]{Chandrasekhar1953}
{Chandrasekhar}, S., \& {Fermi}, E. 1953, \apj, 118, 113,
  \dodoi{10.1086/145731}

\bibitem[{{Chawner} {et~al.}(2020){Chawner}, {Gomez}, {Matsuura}, {Smith},
  {Papageorgiou}, {Rho}, {Noriega-Crespo}, {De Looze}, {Barlow}, {Cigan},
  {Dunne}, \& {Marsh}}]{Chawner20}
{Chawner}, H., {Gomez}, H.~L., {Matsuura}, M., {et~al.} 2020, \mnras, 493,
  2706, \dodoi{10.1093/mnras/staa221}

\bibitem[{{Chen} {et~al.}(2017){Chen}, {Jiang}, {Tamura}, {Kwon}, \&
  {Roman-Lopes}}]{chen17}
{Chen}, Z., {Jiang}, Z., {Tamura}, M., {Kwon}, J., \& {Roman-Lopes}, A. 2017,
  \apj, 838, 80, \dodoi{10.3847/1538-4357/aa65d3}

\bibitem[{{Churchwell} {et~al.}(2009){Churchwell}, {Babler}, {Meade},
  {Whitney}, {Benjamin}, {Indebetouw}, {Cyganowski}, {Robitaille}, {Povich},
  {Watson}, \& {Bracker}}]{Churchwell09}
{Churchwell}, E., {Babler}, B.~L., {Meade}, M.~R., {et~al.} 2009, \pasp, 121,
  213, \dodoi{10.1086/597811}

\bibitem[{{Chuss} {et~al.}(2019){Chuss}, {Andersson}, {Bally}, {Dotson},
  {Dowell}, {Guerra}, {Harper}, {Houde}, {Jones}, {Lazarian}, {Lopez
  Rodriguez}, {Michail}, {Morris}, {Novak}, {Siah}, {Staguhn}, {Vaillancourt},
  {Volpert}, {Werner}, {Wollack}, {Benford}, {Berthoud}, {Cox}, {Crutcher},
  {Dale}, {Fissel}, {Goldsmith}, {Hamilton}, {Hanany}, {Henning}, {Looney},
  {Moseley}, {Santos}, {Stephens}, {Tassis}, {Trinh}, {Van Camp},
  {Ward-Thompson}, \& {HAWC + Science Team}}]{Chuss2019}
{Chuss}, D.~T., {Andersson}, B.~G., {Bally}, J., {et~al.} 2019, \apj, 872, 187,
  \dodoi{10.3847/1538-4357/aafd37}

\bibitem[{{Cosentino} {et~al.}(2022){Cosentino}, {Jim{\'e}nez-Serra}, {Tan},
  {Henshaw}, {Barnes}, {Law}, {Zeng}, {Fontani}, {Caselli}, {Viti}, {Zahorecz},
  {Rico-Villas}, {Meg{\'\i}as}, {Miceli}, {Orlando}, {Ustamujic}, {Greco},
  {Peres}, {Bocchino}, {Fedriani}, {Gorai}, {Testi}, \&
  {Mart{\'\i}n-Pintado}}]{Cosentino22}
{Cosentino}, G., {Jim{\'e}nez-Serra}, I., {Tan}, J.~C., {et~al.} 2022, \mnras,
  511, 953, \dodoi{10.1093/mnras/stac070}

\bibitem[{{Davis}(1951)}]{Davis1951}
{Davis}, L. 1951, Physical Review, 81, 890, \dodoi{10.1103/PhysRev.81.890.2}

\bibitem[{{Deharveng} {et~al.}(2010){Deharveng}, {Schuller}, {Anderson},
  {Zavagno}, {Wyrowski}, {Menten}, {Bronfman}, {Testi}, {Walmsley}, \&
  {Wienen}}]{Deharveng10}
{Deharveng}, L., {Schuller}, F., {Anderson}, L.~D., {et~al.} 2010, \aap, 523,
  A6, \dodoi{10.1051/0004-6361/201014422}

\bibitem[{{Egan} {et~al.}(1998){Egan}, {Shipman}, {Price}, {Carey}, {Clark}, \&
  {Cohen}}]{Egan98}
{Egan}, M.~P., {Shipman}, R.~F., {Price}, S.~D., {et~al.} 1998, \apjl, 494,
  L199, \dodoi{10.1086/311198}

\bibitem[{{Elmegreen} \& {Scalo}(2004)}]{Elmegreen2004}
{Elmegreen}, B.~G., \& {Scalo}, J. 2004, \araa, 42, 211,
  \dodoi{10.1146/annurev.astro.41.011802.094859}

\bibitem[{{Ferri{\`e}re}(2009)}]{Ferriere09}
{Ferri{\`e}re}, K. 2009, \aap, 505, 1183, \dodoi{10.1051/0004-6361/200912617}

\bibitem[{Foreman-Mackey(2016)}]{Foreman-Mackey2016}
Foreman-Mackey, D. 2016, JOSS, 24, \dodoi{10.21105/joss.00024}

\bibitem[{Foreman-Mackey {et~al.}(2013)Foreman-Mackey, Hogg, Lang, \&
  Goodman}]{Foreman-Mackey2013}
Foreman-Mackey, D., Hogg, D.~W., Lang, D., \& Goodman, J. 2013, PASP, 125, 306.
\newblock \url{http://stacks.iop.org/1538-3873/125/i=925/a=306}

\bibitem[{{Ginsburg} {et~al.}(2016){Ginsburg}, {Henkel}, {Ao}, {Riquelme},
  {Kauffmann}, {Pillai}, {Mills}, {Requena-Torres}, {Immer}, {Testi}, {Ott},
  {Bally}, {Battersby}, {Darling}, {Aalto}, {Stanke}, {Kendrew}, {Kruijssen},
  {Longmore}, {Dale}, {Guesten}, \& {Menten}}]{ginsburg16}
{Ginsburg}, A., {Henkel}, C., {Ao}, Y., {et~al.} 2016, \aap, 586, A50,
  \dodoi{10.1051/0004-6361/201526100}

\bibitem[{{Ginsburg} {et~al.}(2020){Ginsburg}, {Anderson}, {Dicker}, {Romero},
  {Svoboda}, {Devlin}, {Galv{\'a}n-Madrid}, {Indebetouw}, {Liu}, {Mason},
  {Mroczkowski}, {Armentrout}, {Bally}, {Brogan}, {Butterfield}, {Hunter},
  {Reese}, {Rosolowsky}, {Sarazin}, {Shirley}, {Sievers}, \&
  {Stanchfield}}]{ginsburg20}
{Ginsburg}, A., {Anderson}, L.~D., {Dicker}, S., {et~al.} 2020, \apjs, 248, 24,
  \dodoi{10.3847/1538-4365/ab8b5c}

\bibitem[{{Gordon} {et~al.}(2018){Gordon}, {Lopez-Rodriguez}, {Andersson},
  {Clarke}, {Coude}, {Moullet}, {Richards}, {Shuping}, {Vacca}, \&
  {Yorke}}]{Gordon18}
{Gordon}, M.~S., {Lopez-Rodriguez}, E., {Andersson}, B.~G., {et~al.} 2018,
  arXiv e-prints, arXiv:1811.03100, \dodoi{10.48550/arXiv.1811.03100}

\bibitem[{{Gravity Collaboration} {et~al.}(2019){Gravity Collaboration},
  {Abuter}, {Amorim}, {Baub{\"o}ck}, {Berger}, {Bonnet}, {Brandner},
  {Cl{\'e}net}, {Coud{\'e} Du Foresto}, {de Zeeuw}, {Dexter}, {Duvert},
  {Eckart}, {Eisenhauer}, {F{\"o}rster Schreiber}, {Garcia}, {Gao}, {Gendron},
  {Genzel}, {Gerhard}, {Gillessen}, {Habibi}, {Haubois}, {Henning}, {Hippler},
  {Horrobin}, {Jim{\'e}nez-Rosales}, {Jocou}, {Kervella}, {Lacour},
  {Lapeyr{\`e}re}, {Le Bouquin}, {L{\'e}na}, {Ott}, {Paumard}, {Perraut},
  {Perrin}, {Pfuhl}, {Rabien}, {Rodriguez Coira}, {Rousset}, {Scheithauer},
  {Sternberg}, {Straub}, {Straubmeier}, {Sturm}, {Tacconi}, {Vincent}, {von
  Fellenberg}, {Waisberg}, {Widmann}, {Wieprecht}, {Wiezorrek}, {Woillez}, \&
  {Yazici}}]{gravity}
{Gravity Collaboration}, {Abuter}, R., {Amorim}, A., {et~al.} 2019, \aap, 625,
  L10, \dodoi{10.1051/0004-6361/201935656}

\bibitem[{{Guerra} {et~al.}(2021){Guerra}, {Chuss}, {Dowell}, {Houde},
  {Michail}, {Siah}, \& {Wollack}}]{Guerra2021}
{Guerra}, J.~A., {Chuss}, D.~T., {Dowell}, C.~D., {et~al.} 2021, \apj, 908, 98,
  \dodoi{10.3847/1538-4357/abd6f0}

\bibitem[{{Guerra} {et~al.}(2023){Guerra}, {Lopez-Rodriguez}, {Chuss},
  {Butterfield}, \& {Schmelz}}]{2023AJ....166...37G}
{Guerra}, J.~A., {Lopez-Rodriguez}, E., {Chuss}, D.~T., {Butterfield}, N.~O.,
  \& {Schmelz}, J.~T. 2023, \aj, 166, 37, \dodoi{10.3847/1538-3881/acdacd}

\bibitem[{{Harper} {et~al.}(2018){Harper}, {Runyan}, {Dowell}, {Wirth},
  {Amato}, {Ames}, {Amiri}, {Banks}, {Bartels}, {Benford}, {Berthoud},
  {Buchanan}, {Casey}, {Chapman}, {Chuss}, {Cook}, {Derro}, {Dotson}, {Evans},
  {Fixsen}, {Gatley}, {Guerra}, {Halpern}, {Hamilton}, {Hamlin}, {Hansen},
  {Heimsath}, {Hermida}, {Hilton}, {Hirsch}, {Hollister}, {Hostetter}, {Irwin},
  {Jhabvala}, {Jhabvala}, {Kastner}, {Kov{\'a}cs}, {Lin}, {Loewenstein},
  {Looney}, {Lopez-Rodriguez}, {Maher}, {Michail}, {Miller}, {Moseley},
  {Novak}, {Pernic}, {Rennick}, {Rhody}, {Sandberg}, {Sandford}, {Santos},
  {Shafer}, {Sharp}, {Shirron}, {Siah}, {Silverberg}, {Sparr}, {Spotz},
  {Staguhn}, {Toorian}, {Towey}, {Tuttle}, {Vaillancourt}, {Voellmer},
  {Volpert}, {Wang}, \& {Wollack}}]{Harper2018}
{Harper}, D.~A., {Runyan}, M.~C., {Dowell}, C.~D., {et~al.} 2018, Journal of
  Astronomical Instrumentation, 7, 1840008, \dodoi{10.1142/S2251171718400081}

\bibitem[{{Hennebelle} \& {Inutsuka}(2019)}]{Hennebelle19}
{Hennebelle}, P., \& {Inutsuka}, S.-i. 2019, Frontiers in Astronomy and Space
  Sciences, 6, 5, \dodoi{10.3389/fspas.2019.00005}

\bibitem[{{Heyer} {et~al.}(2009){Heyer}, {Krawczyk}, {Duval}, \&
  {Jackson}}]{Heyer2009}
{Heyer}, M., {Krawczyk}, C., {Duval}, J., \& {Jackson}, J.~M. 2009, \apj, 699,
  1092, \dodoi{10.1088/0004-637X/699/2/1092}

\bibitem[{{Heywood} {et~al.}(2022){Heywood}, {Rammala}, {Camilo}, {Cotton},
  {Yusef-Zadeh}, {Abbott}, {Adam}, {Adams}, {Aldera}, {Asad}, {Bauermeister},
  {Bennett}, {Bester}, {Bode}, {Botha}, {Botha}, {Brederode}, {Buchner},
  {Burger}, {Cheetham}, {de Villiers}, {Dikgale-Mahlakoana}, {du Toit},
  {Esterhuyse}, {Fanaroff}, {February}, {Fourie}, {Frank}, {Gamatham}, {Geyer},
  {Goedhart}, {Gouws}, {Gumede}, {Hlakola}, {Hokwana}, {Hoosen}, {Horrell},
  {Hugo}, {Isaacson}, {J{\'o}zsa}, {Jonas}, {Joubert}, {Julie}, {Kapp},
  {Kenyon}, {Kotz{\'e}}, {Kriek}, {Kriel}, {Krishnan}, {Lehmensiek},
  {Liebenberg}, {Lord}, {Lunsky}, {Madisa}, {Magnus}, {Mahgoub}, {Makhaba},
  {Makhathini}, {Malan}, {Manley}, {Marais}, {Martens}, {Mauch}, {Merry},
  {Millenaar}, {Mnyandu}, {Mokone}, {Monama}, {Mphego}, {New}, {Ngcebetsha},
  {Ngoasheng}, {Ockards}, {Oozeer}, {Otto}, {Passmoor}, {Patel}, {Peens-Hough},
  {Perkins}, {Ramaila}, {Ramanujam}, {Ramudzuli}, {Ratcliffe}, {Robyntjies},
  {Salie}, {Sambu}, {Schollar}, {Schwardt}, {Schwartz}, {Serylak}, {Siebrits},
  {Sirothia}, {Slabber}, {Smirnov}, {Sofeya}, {Taljaard}, {Tasse}, {Tiplady},
  {Toruvanda}, {Twum}, {van Balla}, {van der Byl}, {van der Merwe}, {Van
  Tonder}, {Van Wyk}, {Venter}, {Venter}, {Wallace}, {Welz}, {Williams}, \&
  {Xaia}}]{heywood22}
{Heywood}, I., {Rammala}, I., {Camilo}, F., {et~al.} 2022, \apj, 925, 165,
  \dodoi{10.3847/1538-4357/ac449a}

\bibitem[{{Hildebrand} {et~al.}(2009){Hildebrand}, {Kirby}, {Dotson}, {Houde},
  \& {Vaillancourt}}]{Hildebrand2009}
{Hildebrand}, R.~H., {Kirby}, L., {Dotson}, J.~L., {Houde}, M., \&
  {Vaillancourt}, J.~E. 2009, \apj, 696, 567,
  \dodoi{10.1088/0004-637X/696/1/567}

\bibitem[{{Houde} {et~al.}(2009){Houde}, {Vaillancourt}, {Hildebrand},
  {Chitsazzadeh}, \& {Kirby}}]{Houde2009}
{Houde}, M., {Vaillancourt}, J.~E., {Hildebrand}, R.~H., {Chitsazzadeh}, S., \&
  {Kirby}, L. 2009, \apj, 706, 1504, \dodoi{10.1088/0004-637X/706/2/1504}

\bibitem[{Hunter(2007)}]{Hunter2007}
Hunter, J.~D. 2007, CSE, 9

\bibitem[{{Inoue} {et~al.}(2009){Inoue}, {Yamazaki}, \& {Inutsuka}}]{Inoue2009}
{Inoue}, T., {Yamazaki}, R., \& {Inutsuka}, S.-i. 2009, \apj, 695, 825,
  \dodoi{10.1088/0004-637X/695/2/825}

\bibitem[{{International Consortium Of Scientists}(2011)}]{2011ascl.soft07013I}
{International Consortium Of Scientists}. 2011, {CASA: Common Astronomy
  Software Applications}, Astrophysics Source Code Library.
\newblock \doeprint{1107.013}

\bibitem[{Jones {et~al.}(2001)Jones, Oliphant, \& et~al.}]{Jones2001}
Jones, E., Oliphant, T., \& et~al., P.~P. 2001, SciPy: Open Source Scientific
  Tools for Python.
\newblock \url{http://www.scipy.org/}

\bibitem[{{Jones} {et~al.}(2012){Jones}, {Burton}, {Cunningham},
  {Requena-Torres}, {Menten}, {Schilke}, {Belloche}, {Leurini},
  {Mart{\'\i}n-Pintado}, {Ott}, \& {Walsh}}]{jones12}
{Jones}, P.~A., {Burton}, M.~G., {Cunningham}, M.~R., {et~al.} 2012, \mnras,
  419, 2961, \dodoi{10.1111/j.1365-2966.2011.19941.x}

\bibitem[{{Kruijssen} {et~al.}(2015){Kruijssen}, {Dale}, \& {Longmore}}]{Kru15}
{Kruijssen}, J.~M.~D., {Dale}, J.~E., \& {Longmore}, S.~N. 2015, \mnras, 447,
  1059, \dodoi{10.1093/mnras/stu2526}

\bibitem[{{LaRosa} {et~al.}(2000){LaRosa}, {Kassim}, {Lazio}, \&
  {Hyman}}]{LaRosa00}
{LaRosa}, T.~N., {Kassim}, N.~E., {Lazio}, T. J.~W., \& {Hyman}, S.~D. 2000,
  \aj, 119, 207, \dodoi{10.1086/301168}

\bibitem[{{Larson}(1981)}]{Larson81}
{Larson}, R.~B. 1981, \mnras, 194, 809, \dodoi{10.1093/mnras/194.4.809}

\bibitem[{Lazarian \& Hoang(2007)}]{Lazarian07}
Lazarian, A., \& Hoang, T. 2007, Monthly Notices of the Royal Astronomical
  Society, 378, 910, \dodoi{10.1111/j.1365-2966.2007.11817.x}

\bibitem[{{Lebigot}(2016)}]{LebigotUnc}
{Lebigot}, E. 2016, \dodoi{http://pythonhosted.org/uncertainties/}

\bibitem[{{Lockman}(1989)}]{Lockman89}
{Lockman}, F.~J. 1989, \apjs, 71, 469, \dodoi{10.1086/191383}

\bibitem[{{Lopez-Rodriguez} {et~al.}(2021){Lopez-Rodriguez}, {Guerra},
  {Asgari-Targhi}, \& {Schmelz}}]{Lopez_Rodriguez2021}
{Lopez-Rodriguez}, E., {Guerra}, J.~A., {Asgari-Targhi}, M., \& {Schmelz},
  J.~T. 2021, \apj, 914, 24, \dodoi{10.3847/1538-4357/abf934}

\bibitem[{{Mangilli} {et~al.}(2019){Mangilli}, {Aumont}, {Bernard}, {Buzzelli},
  {de Gasperis}, {Durrive}, {Ferriere}, {Fo{\"e}nard}, {Hughes}, {Lacourt},
  {Misawa}, {Montier}, {Mot}, {Ristorcelli}, {Roussel}, {Ade}, {Alina}, {de
  Bernardis}, {de Gouveia Dal Pino}, {Dubois}, {Engel}, {Guillet}, {Hargrave},
  {Laureijs}, {Longval}, {Maffei}, {Magalhaes}, {Marty}, {Masi}, {Montel},
  {Pajot}, {P{\'e}rot}, {Rodriguez}, {Salatino}, {Saccoccio}, {Savini},
  {Stever}, {Tauber}, {Tibbs}, \& {Tucker}}]{Mangilli19}
{Mangilli}, A., {Aumont}, J., {Bernard}, J.~P., {et~al.} 2019, \aap, 630, A74,
  \dodoi{10.1051/0004-6361/201935072}

\bibitem[{{Marsh} {et~al.}(2015){Marsh}, {Whitworth}, \& {Lomax}}]{Marsh2015}
{Marsh}, K.~A., {Whitworth}, A.~P., \& {Lomax}, O. 2015, \mnras, 454, 4282,
  \dodoi{10.1093/mnras/stv2248}

\bibitem[{{Marsh} {et~al.}(2017){Marsh}, {Whitworth}, {Lomax}, {Ragan},
  {Becciani}, {Cambr{\'e}sy}, {Di Giorgio}, {Eden}, {Elia}, {Kacsuk},
  {Molinari}, {Palmeirim}, {Pezzuto}, {Schneider}, {Sciacca}, \&
  {Vitello}}]{Marsh2017}
{Marsh}, K.~A., {Whitworth}, A.~P., {Lomax}, O., {et~al.} 2017, \mnras, 471,
  2730, \dodoi{10.1093/mnras/stx1723}

\bibitem[{{Matsuura} {et~al.}(2017){Matsuura}, {Indebetouw}, {Woosley},
  {Bujarrabal}, {Abell{\'a}n}, {McCray}, {Kamenetzky}, {Fransson}, {Barlow},
  {Gomez}, {Cigan}, {De Looze}, {Spyromilio}, {Staveley-Smith}, {Zanardo},
  {Roche}, {Larsson}, {Viti}, {van Loon}, {Wheeler}, {Baes}, {Chevalier},
  {Lundqvist}, {Marcaide}, {Dwek}, {Meixner}, {Ng}, {Sonneborn}, \&
  {Yates}}]{Matsuura17}
{Matsuura}, M., {Indebetouw}, R., {Woosley}, S., {et~al.} 2017, \mnras, 469,
  3347, \dodoi{10.1093/mnras/stx830}

\bibitem[{{McClure-Griffiths} {et~al.}(2009){McClure-Griffiths}, {Pisano},
  {Calabretta}, {Ford}, {Lockman}, {Staveley-Smith}, {Kalberla}, {Bailin},
  {Dedes}, {Janowiecki}, {Gibson}, {Murphy}, {Nakanishi}, \&
  {Newton-McGee}}]{McClure-Griffiths09}
{McClure-Griffiths}, N.~M., {Pisano}, D.~J., {Calabretta}, M.~R., {et~al.}
  2009, \apjs, 181, 398, \dodoi{10.1088/0067-0049/181/2/398}

\bibitem[{{McMullin} {et~al.}(2007){McMullin}, {Waters}, {Schiebel}, {Young},
  \& {Golap}}]{Casa}
{McMullin}, J.~P., {Waters}, B., {Schiebel}, D., {Young}, W., \& {Golap}, K.
  2007, in Astronomical Society of the Pacific Conference Series, Vol. 376,
  Astronomical Data Analysis Software and Systems XVI, ed. R.~A. {Shaw},
  F.~{Hill}, \& D.~J. {Bell}, 127

\bibitem[{{Mestel}(1966)}]{Mestel1966}
{Mestel}, L. 1966, \mnras, 133, 265, \dodoi{10.1093/mnras/133.2.265}

\bibitem[{{Mills} \& {Battersby}(2017)}]{mills17}
{Mills}, E.~A.~C., \& {Battersby}, C. 2017, \apj, 835, 76,
  \dodoi{10.3847/1538-4357/835/1/76}

\bibitem[{{Milne}(1990)}]{milne90}
{Milne}, D.~K. 1990, in Galactic and Intergalactic Magnetic Fields, ed.
  R.~{Beck}, P.~P. {Kronberg}, \& R.~{Wielebinski}, Vol. 140, 67

\bibitem[{{Molinari} {et~al.}(2010){Molinari}, {Swinyard}, {Bally}, {Barlow},
  {Bernard}, {Martin}, {Moore}, {Noriega-Crespo}, {Plume}, {Testi}, {Zavagno},
  {Abergel}, {Ali}, {Andr{\'e}}, {Baluteau}, {Benedettini}, {Bern{\'e}},
  {Billot}, {Blommaert}, {Bontemps}, {Boulanger}, {Brand}, {Brunt}, {Burton},
  {Campeggio}, {Carey}, {Caselli}, {Cesaroni}, {Cernicharo}, {Chakrabarti},
  {Chrysostomou}, {Codella}, {Cohen}, {Compiegne}, {Davis}, {de Bernardis}, {de
  Gasperis}, {Di Francesco}, {di Giorgio}, {Elia}, {Faustini}, {Fischera},
  {Fukui}, {Fuller}, {Ganga}, {Garcia-Lario}, {Giard}, {Giardino}, {Glenn},
  {Goldsmith}, {Griffin}, {Hoare}, {Huang}, {Jiang}, {Joblin}, {Joncas},
  {Juvela}, {Kirk}, {Lagache}, {Li}, {Lim}, {Lord}, {Lucas}, {Maiolo},
  {Marengo}, {Marshall}, {Masi}, {Massi}, {Matsuura}, {Meny}, {Minier},
  {Miville-Desch{\^e}nes}, {Montier}, {Motte}, {M{\"u}ller}, {Natoli}, {Neves},
  {Olmi}, {Paladini}, {Paradis}, {Pestalozzi}, {Pezzuto}, {Piacentini},
  {Pomar{\`e}s}, {Popescu}, {Reach}, {Richer}, {Ristorcelli}, {Roy}, {Royer},
  {Russeil}, {Saraceno}, {Sauvage}, {Schilke}, {Schneider-Bontemps},
  {Schuller}, {Schultz}, {Shepherd}, {Sibthorpe}, {Smith}, {Smith},
  {Spinoglio}, {Stamatellos}, {Strafella}, {Stringfellow}, {Sturm}, {Taylor},
  {Thompson}, {Tuffs}, {Umana}, {Valenziano}, {Vavrek}, {Viti}, {Waelkens},
  {Ward-Thompson}, {White}, {Wyrowski}, {Yorke}, \& {Zhang}}]{molinari10}
{Molinari}, S., {Swinyard}, B., {Bally}, J., {et~al.} 2010, \pasp, 122, 314,
  \dodoi{10.1086/651314}

\bibitem[{{Molinari} {et~al.}(2011){Molinari}, {Bally}, {Noriega-Crespo},
  {Compi{\`e}gne}, {Bernard}, {Paradis}, {Martin}, {Testi}, {Barlow}, {Moore},
  {Plume}, {Swinyard}, {Zavagno}, {Calzoletti}, {Di Giorgio}, {Elia},
  {Faustini}, {Natoli}, {Pestalozzi}, {Pezzuto}, {Piacentini}, {Polenta},
  {Polychroni}, {Schisano}, {Traficante}, {Veneziani}, {Battersby}, {Burton},
  {Carey}, {Fukui}, {Li}, {Lord}, {Morgan}, {Motte}, {Schuller},
  {Stringfellow}, {Tan}, {Thompson}, {Ward-Thompson}, {White}, \&
  {Umana}}]{molinari11}
{Molinari}, S., {Bally}, J., {Noriega-Crespo}, A., {et~al.} 2011, \apjl, 735,
  L33, \dodoi{10.1088/2041-8205/735/2/L33}

\bibitem[{{Molinari} {et~al.}(2016){Molinari}, {Schisano}, {Elia},
  {Pestalozzi}, {Traficante}, {Pezzuto}, {Swinyard}, {Noriega-Crespo}, {Bally},
  {Moore}, {Plume}, {Zavagno}, {di Giorgio}, {Liu}, {Pilbratt}, {Mottram},
  {Russeil}, {Piazzo}, {Veneziani}, {Benedettini}, {Calzoletti}, {Faustini},
  {Natoli}, {Piacentini}, {Merello}, {Palmese}, {Del Grande}, {Polychroni},
  {Rygl}, {Polenta}, {Barlow}, {Bernard}, {Martin}, {Testi}, {Ali},
  {Andr{\'e}}, {Beltr{\'a}n}, {Billot}, {Carey}, {Cesaroni}, {Compi{\`e}gne},
  {Eden}, {Fukui}, {Garcia-Lario}, {Hoare}, {Huang}, {Joncas}, {Lim}, {Lord},
  {Martinavarro-Armengol}, {Motte}, {Paladini}, {Paradis}, {Peretto},
  {Robitaille}, {Schilke}, {Schneider}, {Schulz}, {Sibthorpe}, {Strafella},
  {Thompson}, {Umana}, {Ward-Thompson}, \& {Wyrowski}}]{molinari16}
{Molinari}, S., {Schisano}, E., {Elia}, D., {et~al.} 2016, \aap, 591, A149,
  \dodoi{10.1051/0004-6361/201526380}

\bibitem[{{Morris} \& {Serabyn}(1996)}]{morris96}
{Morris}, M., \& {Serabyn}, E. 1996, \araa, 34, 645,
  \dodoi{10.1146/annurev.astro.34.1.645}

\bibitem[{{Oka} {et~al.}(2001){Oka}, {Hasegawa}, {Sato}, {Tsuboi}, \&
  {Miyazaki}}]{oka01}
{Oka}, T., {Hasegawa}, T., {Sato}, F., {Tsuboi}, M., \& {Miyazaki}, A. 2001,
  \pasj, 53, 787, \dodoi{10.1093/pasj/53.5.787}

\bibitem[{{Omont} {et~al.}(2003){Omont}, {Gilmore}, {Alard}, {Aracil},
  {August}, {Baliyan}, {Beaulieu}, {B{\'e}gon}, {Bertou}, {Blommaert},
  {Borsenberger}, {Burgdorf}, {Caillaud}, {Cesarsky}, {Chitre}, {Copet}, {de
  Batz}, {Egan}, {Egret}, {Epchtein}, {Felli}, {Fouqu{\'e}}, {Ganesh},
  {Genzel}, {Glass}, {Gredel}, {Groenewegen}, {Guglielmo}, {Habing},
  {Hennebelle}, {Jiang}, {Joshi}, {Kimeswenger}, {Messineo},
  {Miville-Desch{\^e}nes}, {Moneti}, {Morris}, {Ojha}, {Ortiz}, {Ott},
  {Parthasarathy}, {P{\'e}rault}, {Price}, {Robin}, {Schultheis}, {Schuller},
  {Simon}, {Soive}, {Testi}, {Teyssier}, {Tiph{\`e}ne}, {Unavane}, {van Loon},
  \& {Wyse}}]{Omont03}
{Omont}, A., {Gilmore}, G.~F., {Alard}, C., {et~al.} 2003, \aap, 403, 975,
  \dodoi{10.1051/0004-6361:20030437}

\bibitem[{{Ostriker} {et~al.}(2001){Ostriker}, {Stone}, \&
  {Gammie}}]{Ostriker2001}
{Ostriker}, E.~C., {Stone}, J.~M., \& {Gammie}, C.~F. 2001, \apj, 546, 980,
  \dodoi{10.1086/318290}

\bibitem[{{Ou} {et~al.}(2018){Ou}, {Chu}, {Maggi}, {Li}, {Chang}, \&
  {Gruendl}}]{Ou18}
{Ou}, P.-S., {Chu}, Y.-H., {Maggi}, P., {et~al.} 2018, \apj, 863, 137,
  \dodoi{10.3847/1538-4357/aad04b}

\bibitem[{{Par{\'e}} {et~al.}(2024){Par{\'e}}, {Butterfield}, {Chuss},
  {Guerra}, {Iuliano}, {Karpovich}, {Morris}, \& {Wollack}}]{Pare24}
{Par{\'e}}, D., {Butterfield}, N.~O., {Chuss}, D.~T., {et~al.} 2024, arXiv
  e-prints, arXiv:2401.05317, \dodoi{10.48550/arXiv.2401.05317}

\bibitem[{{Parsons} {et~al.}(2018){Parsons}, {Dempsey}, {Thomas}, {Berry},
  {Currie}, {Friberg}, {Wouterloot}, {Chrysostomou}, {Graves}, {Tilanus},
  {Bell}, \& {Rawlings}}]{parsons18}
{Parsons}, H., {Dempsey}, J.~T., {Thomas}, H.~S., {et~al.} 2018, \apjs, 234,
  22, \dodoi{10.3847/1538-4365/aa989c}

\bibitem[{P\'{e}rez \& Granger(2007)}]{Perez2007}
P\'{e}rez, F., \& Granger, B.~E. 2007, CSE, 9, \dodoi{10.1109/MCSE.2007.53}

\bibitem[{{Pierce-Price} {et~al.}(2000){Pierce-Price}, {Richer}, {Greaves},
  {Holland}, {Jenness}, {Lasenby}, {White}, {Matthews}, {Ward-Thompson},
  {Dent}, {Zylka}, {Mezger}, {Hasegawa}, {Oka}, {Omont}, \&
  {Gilmore}}]{Pierce-Price00}
{Pierce-Price}, D., {Richer}, J.~S., {Greaves}, J.~S., {et~al.} 2000, \apjl,
  545, L121, \dodoi{10.1086/317884}

\bibitem[{{Planck Collaboration} {et~al.}(2011){Planck Collaboration},
  {Abergel}, {Ade}, {Aghanim}, {Arnaud}, {Ashdown}, {Aumont}, {Baccigalupi},
  {Balbi}, {Banday}, {Barreiro}, {Bartlett}, {Battaner}, {Benabed},
  {Beno{\^\i}t}, {Bernard}, {Bersanelli}, {Bhatia}, {Bock}, {Bonaldi}, {Bond},
  {Borrill}, {Bouchet}, {Boulanger}, {Bucher}, {Burigana}, {Cabella},
  {Cardoso}, {Catalano}, {Cay{\'o}n}, {Challinor}, {Chamballu}, {Chiang},
  {Chiang}, {Christensen}, {Clements}, {Colombi}, {Couchot}, {Coulais},
  {Crill}, {Cuttaia}, {Danese}, {Davies}, {Davis}, {de Bernardis}, {de
  Gasperis}, {de Rosa}, {de Zotti}, {Delabrouille}, {Delouis}, {D{\'e}sert},
  {Dickinson}, {Dobashi}, {Donzelli}, {Dor{\'e}}, {D{\"o}rl}, {Douspis},
  {Dupac}, {Efstathiou}, {En{\ss}lin}, {Eriksen}, {Finelli}, {Forni},
  {Frailis}, {Franceschi}, {Galeotta}, {Ganga}, {Giard}, {Giardino},
  {Giraud-H{\'e}raud}, {Gonz{\'a}lez-Nuevo}, {G{\'o}rski}, {Gratton},
  {Gregorio}, {Gruppuso}, {Guillet}, {Hansen}, {Harrison},
  {Henrot-Versill{\'e}}, {Herranz}, {Hildebrandt}, {Hivon}, {Hobson}, {Holmes},
  {Hovest}, {Hoyland}, {Huffenberger}, {Jaffe}, {Jones}, {Jones}, {Juvela},
  {Keih{\"a}nen}, {Keskitalo}, {Kisner}, {Kneissl}, {Knox}, {Kurki-Suonio},
  {Lagache}, {Lamarre}, {Lasenby}, {Laureijs}, {Lawrence}, {Leach}, {Leonardi},
  {Leroy}, {Linden-V{\o}rnle}, {L{\'o}pez-Caniego}, {Lubin},
  {Mac{\'\i}as-P{\'e}rez}, {MacTavish}, {Maffei}, {Mandolesi}, {Mann}, {Maris},
  {Marshall}, {Martin}, {Mart{\'\i}nez-Gonz{\'a}lez}, {Masi}, {Matarrese},
  {Matthai}, {Mazzotta}, {McGehee}, {Meinhold}, {Melchiorri}, {Mendes},
  {Mennella}, {Mitra}, {Miville-Desch{\^e}nes}, {Moneti}, {Montier},
  {Morgante}, {Mortlock}, {Munshi}, {Murphy}, {Naselsky}, {Natoli},
  {Netterfield}, {N{\o}rgaard-Nielsen}, {Noviello}, {Novikov}, {Novikov},
  {Osborne}, {Pajot}, {Paladini}, {Pasian}, {Patanchon}, {Perdereau},
  {Perotto}, {Perrotta}, {Piacentini}, {Piat}, {Plaszczynski}, {Pointecouteau},
  {Polenta}, {Ponthieu}, {Poutanen}, {Pr{\'e}zeau}, {Prunet}, {Puget}, {Reach},
  {Rebolo}, {Reinecke}, {Renault}, {Ricciardi}, {Riller}, {Ristorcelli},
  {Rocha}, {Rosset}, {Rubi{\~n}o-Mart{\'\i}n}, {Rusholme}, {Sandri}, {Santos},
  {Savini}, {Scott}, {Seiffert}, {Shellard}, {Smoot}, {Starck}, {Stivoli},
  {Stolyarov}, {Sudiwala}, {Sygnet}, {Tauber}, {Terenzi}, {Toffolatti},
  {Tomasi}, {Torre}, {Tristram}, {Tuovinen}, {Umana}, {Valenziano},
  {Verstraete}, {Vielva}, {Villa}, {Vittorio}, {Wade}, {Wandelt}, {Yvon},
  {Zacchei}, \& {Zonca}}]{PlanckCollaboration2011}
{Planck Collaboration}, {Abergel}, A., {Ade}, P.~A.~R., {et~al.} 2011, \aap,
  536, A25, \dodoi{10.1051/0004-6361/201116483}

\bibitem[{{Schuller} {et~al.}(2009){Schuller}, {Menten}, {Contreras},
  {Wyrowski}, {Schilke}, {Bronfman}, {Henning}, {Walmsley}, {Beuther},
  {Bontemps}, {Cesaroni}, {Deharveng}, {Garay}, {Herpin}, {Lefloch}, {Linz},
  {Mardones}, {Minier}, {Molinari}, {Motte}, {Nyman}, {Reveret}, {Risacher},
  {Russeil}, {Schneider}, {Testi}, {Troost}, {Vasyunina}, {Wienen}, {Zavagno},
  {Kovacs}, {Kreysa}, {Siringo}, \& {Wei{\ss}}}]{atlasgal}
{Schuller}, F., {Menten}, K.~M., {Contreras}, Y., {et~al.} 2009, \aap, 504,
  415, \dodoi{10.1051/0004-6361/200811568}

\bibitem[{{Skalidis} \& {Tassis}(2021)}]{Skalidis2021}
{Skalidis}, R., \& {Tassis}, K. 2021, \aap, 647, A186,
  \dodoi{10.1051/0004-6361/202039779}

\bibitem[{{Skrutskie} {et~al.}(2006){Skrutskie}, {Cutri}, {Stiening},
  {Weinberg}, {Schneider}, {Carpenter}, {Beichman}, {Capps}, {Chester},
  {Elias}, {Huchra}, {Liebert}, {Lonsdale}, {Monet}, {Price}, {Seitzer},
  {Jarrett}, {Kirkpatrick}, {Gizis}, {Howard}, {Evans}, {Fowler}, {Fullmer},
  {Hurt}, {Light}, {Kopan}, {Marsh}, {McCallon}, {Tam}, {Van Dyk}, \&
  {Wheelock}}]{2mass}
{Skrutskie}, M.~F., {Cutri}, R.~M., {Stiening}, R., {et~al.} 2006, \aj, 131,
  1163, \dodoi{10.1086/498708}

\bibitem[{{Tsuboi} {et~al.}(2009){Tsuboi}, {Miyazaki}, \& {Okumura}}]{tsuboi09}
{Tsuboi}, M., {Miyazaki}, A., \& {Okumura}, S.~K. 2009, \pasj, 61, 29,
  \dodoi{10.1093/pasj/61.1.29}

\bibitem[{{Tsuboi} {et~al.}(2015){Tsuboi}, {Miyazaki}, \& {Uehara}}]{tsuboi15}
{Tsuboi}, M., {Miyazaki}, A., \& {Uehara}, K. 2015, \pasj, 67, 90,
  \dodoi{10.1093/pasj/psv058}

\bibitem[{{Tsuboi} {et~al.}(1997){Tsuboi}, {Ukita}, \& {Handa}}]{tsuboi97}
{Tsuboi}, M., {Ukita}, N., \& {Handa}, T. 1997, \apj, 481, 263,
  \dodoi{10.1086/304053}

\bibitem[{{van der Walt} {et~al.}(2011){van der Walt}, Colbert, \&
  Varoquaux}]{vanderWalt2011}
{van der Walt}, S., Colbert, S.~C., \& Varoquaux, G. 2011, CSE, 13,
  \dodoi{10.1109/MCSE.2011.37}

\bibitem[{{Vink}(2017)}]{Vink17}
{Vink}, J. 2017, in Handbook of Supernovae, ed. A.~W. {Alsabti} \& P.~{Murdin},
  2063, \dodoi{10.1007/978-3-319-21846-5_92}

\bibitem[{{Vrba} {et~al.}(1987){Vrba}, {Baierlein}, \& {Herbst}}]{Vrba87}
{Vrba}, F.~J., {Baierlein}, R., \& {Herbst}, W. 1987, \apj, 317, 207,
  \dodoi{10.1086/165269}

\bibitem[{{Wang}(2021)}]{wang21}
{Wang}, Q.~D. 2021, \mnras, 504, 1609, \dodoi{10.1093/mnras/stab801}

\bibitem[{{Wang} {et~al.}(2022){Wang}, {Jiang}, {Li}, {Zhao}, \&
  {Ren}}]{wang22}
{Wang}, Y., {Jiang}, B., {Li}, J., {Zhao}, H., \& {Ren}, Y. 2022, \aj, 163, 60,
  \dodoi{10.3847/1538-3881/ac35db}

\bibitem[{{White} {et~al.}(2005){White}, {Becker}, \& {Helfand}}]{White05}
{White}, R.~L., {Becker}, R.~H., \& {Helfand}, D.~J. 2005, \aj, 130, 586,
  \dodoi{10.1086/431249}

\bibitem[{{Wright} {et~al.}(2010){Wright}, {Eisenhardt}, {Mainzer}, {Ressler},
  {Cutri}, {Jarrett}, {Kirkpatrick}, {Padgett}, {McMillan}, {Skrutskie},
  {Stanford}, {Cohen}, {Walker}, {Mather}, {Leisawitz}, {Gautier}, {McLean},
  {Benford}, {Lonsdale}, {Blain}, {Mendez}, {Irace}, {Duval}, {Liu}, {Royer},
  {Heinrichsen}, {Howard}, {Shannon}, {Kendall}, {Walsh}, {Larsen}, {Cardon},
  {Schick}, {Schwalm}, {Abid}, {Fabinsky}, {Naes}, \& {Tsai}}]{Wright10}
{Wright}, E.~L., {Eisenhardt}, P. R.~M., {Mainzer}, A.~K., {et~al.} 2010, \aj,
  140, 1868, \dodoi{10.1088/0004-6256/140/6/1868}

\bibitem[{{Yoast-Hull} {et~al.}(2013){Yoast-Hull}, {Everett}, {Gallagher}, \&
  {Zweibel}}]{Yoast-Hull13}
{Yoast-Hull}, T.~M., {Everett}, J.~E., {Gallagher}, J.~S., I., \& {Zweibel},
  E.~G. 2013, \apj, 768, 53, \dodoi{10.1088/0004-637X/768/1/53}

\bibitem[{{Yu} {et~al.}(2018){Yu}, {Xu}, \& {Wang}}]{Yu18}
{Yu}, N.-P., {Xu}, J.-L., \& {Wang}, J.-J. 2018, Research in Astronomy and
  Astrophysics, 18, 015, \dodoi{10.1088/1674-4527/18/2/15}

\end{thebibliography}

\end{document}